\documentstyle[preprint,epsfig,aps]{revtex}   
\newcommand{\nc}{\newcommand}
\nc{\be}{\begin{equation}}
\nc{\ee}{\end{equation}}
\nc{\bea}{\begin{eqnarray}}
\nc{\eea}{\end{eqnarray}}
\nc{\beas}{\begin{eqnarray*}}
\nc{\eeas}{\end{eqnarray*}}
\nc{\noi}{\noindent}
\nc{\sD}{\not \! \! D}
\nc{\s}[1]{\not \! #1}
\nc{\non}{\nonumber}
\nc{\bb}{\bibitem}
\nc{\lf}{\left}
\nc{\ri}{\right}
\nc{\mb}[1]{\makebox[#1]{}}
\nc{\pa}{\partial}
\nc{\sA}{\not \! \! A}
\nc{\newsec}[1]{\section{#1}\mb{0.5cm}}
\nc{\h}{\frac{1}{2}}
\nc{\ra}{\rightarrow}
\nc{\la}{\leftarrow}
\nc{\ep}{$e^+e^-\ra\pi^+\pi^-\;$}
\nc{\emuon}{$e^+e^-\ra\mu^+\mu^-\;$}
\nc{\epp}{$e^+e^-\ra\pi^+\pi^0\pi^-\;$}
\nc{\elec}{$e^+e^-\ra\gamma^*\ra e^+e^-\;$}
\nc{\omg}{\omega}
\nc{\mor}{\omega \rho}
\def\mathunderaccent#1{\let\theaccent#1\mathpalette\putaccentunder}
\def\putaccentunder#1#2{\oalign{$#1#2$\crcr\hidewidth
\vbox to.2ex{\hbox{$#1\theaccent{}$}\vss}\hidewidth}}

\nc{\ti}{\mathunderaccent\tilde}
\nc{\M}{{\cal M}}
\nc{\rw}{$\rho\!-\!\omega\;$}

\def\hhhb{\rule[-3.mm]{0.mm}{9.mm}}
\def\hhhc{\rule[-3.mm]{0.mm}{3.mm}}

\begin{document}
\tightenlines    
\draft           
\preprint{\vbox{~~~ \\
                                       \null \hfill LPNHE 2002--10 \\
                                       \null \hfill FERMILAB-PUB-02-294 \\
                                       \null\hfill nucl-th/0301037 }}
\title{The Pion Form  Factor\\
Within the Hidden Local Symmetry Model}
\author{M.~Benayoun$(^{a,b})$, P. David$(^{b})$, L. DelBuono$(^{b})$,  
        Ph. Leruste$(^{b})$, H.B. O'Connell$(^{c})$\\ 
}
\address{(~$^a$) CERN, Laboratoire Europ\'een pour la Recherche Nucl\'eaire,
1211, Gen\`eve 23, Switzerland \\ 
(~$^b$) LPNHE des Universit\'es Paris VI et VII--IN2P3, Paris,
         France      \\    
(~$^c$)	 Fermilab, PO Box 500 MS 109, Batavia IL 60510, USA.}
\date{6th December 2002}
\maketitle
\begin{abstract}
We analyze a pion form factor formulation which fulfills
the Analyticity requirement within the Hidden Local Symmetry
(HLS) Model. This implies an $s$--dependent dressing 
of the $\rho-\gamma$ VMD coupling and an account of 
several coupled channels. The corresponding function $F_\pi(s)$
provides nice fits of the pion form factor data from
$s=-0.25$ to $s=1$ GeV$^2$. It is shown that the coupling to 
$K \overline{K}$ has little effect, while $\omg \pi^0$ improves 
significantly the fit probability below the $\phi$ mass. 
No need for additional states like $\rho(1450)$ shows up in this
invariant--mass range.
All parameters, except for the subtraction polynomial coefficients,
are fixed from the rest of the HLS phenomenology. The fits show
consistency with the expected behaviour of $F_\pi(s)$ at $s=0$
up to ${\cal O} (s^2)$ and with the phase shift data on  
$\delta_1^1(s)$ from threshold to somewhat above the $\phi$ mass. 
The $\omg$  sector is also examined in relation with recent 
data from CMD--2.

\end{abstract}

\newpage

\pagenumbering{arabic}
\section{Introduction}
\label{One}

\indent
In the physics of exclusive processes, the pion form factor $F_\pi(s)$
plays an important role. It is indeed a fundamental tool
in order to estimate precisely the hadronic contribution to the muon 
anomalous magnetic moment (for recent works, see \cite{Yndurain}
and \cite{Davier} where an exhaustive list of references can be 
found). It is also an important information, as it allows to
test the predictions of Chiral perturbation theory (ChPT) 
which describes the behaviour of QCD at low energies
where non--perturbative effects dominate.
Among very recent works on this classical subject, let us quote 
Refs. \cite{Yndurain,Pich1,Pich2}.

Several descriptions of the pion form factor are  proposed.
For instance, Ref. \cite{Yndurain} gives a parametrization of 
the $P-$wave 
$\pi \pi$ phase shift $\delta_1^1(s)$ derived from general 
analyticity principles supplemented with some properties related 
with the existence of the $\rho^0(770)$ meson. Watson theorem
relates $F_\pi(s)$ with the $\pi \pi$ phase shift  by proving
that $\rm{Arg[F}_\pi(s)] = \delta_1^1(s)$ up to the first
inelastic threshold. In principle, this is located at the four--pion
threshold, however experimental data \cite{Ochs},
especially on $P$--wave inelasticity, show that
$ \delta_1^1(s)$ can be considered elastic with a nice precision
up to the 0.95 GeV region. The free parameters of the function
defined by \cite{Yndurain} are fitted on Aleph \cite{Aleph} and Opal 
\cite{Opal} $\tau$ decay data on the two--pion final state. 
The derived phase \cite{Yndurain} is shown to predict
impressively the phase of Ref. \cite{Protopescu}. 
In this approach, the role of the $\rho(770)$
meson is obvious~; what is less obvious is whether
additional states like the $\rho(1450)$ play any role
below $\sqrt{s} =1$ GeV.
Actually, while focussing on estimating hadronic
contributions to the muon anomalous magnetic moment,
it is not a real concern.

In the same spirit, Ref. \cite{Pich1} starts from phase 
shift data \cite{Ochs} measured up to $\sqrt{s} \simeq 2$ GeV, 
assumes Watson theorem and fit the Aleph \cite{Aleph} 
and CLEOII \cite{Cleo} relevant data sets with~:

$$
F_\pi(s)=\exp{\left \{
\alpha_1 s + \frac{1}{2}\alpha_2 s^2+ \frac{s^3}{\pi}
\int_{4 m_\pi^2}^{\Lambda^2} \frac{dz}{z^3} 
\frac{\delta_1^1(z)}{z-s-i\epsilon}
\right \}
}
$$

\noindent where $\Lambda$ is some cut--off and $\alpha_1$ and 
$\alpha_2$ are free parameters.

The approach of Ref. \cite{Pich2} relies instead on the Resonance 
Chiral Theory developed in \cite{Ecker}, where vector mesons are
explicitly introduced in the Lagrangian. Here the parameters 
to be fitted are the masses and couplings associated with the usual
vector meson nonet (those containing the $\rho(770)$) and
the one associated with  the $\rho(1450)$ meson. Focusing on the 
$\rho(770)$ nonet, this mass is fit as $M_{V_1} \simeq 840$ MeV, 
which does not prevent the Breit--Wigner $\rho(770)$ parameters 
derived from this fit \cite{Pich2}  to be very close to expectations \cite{PDG02}.
Here again, the phase predicted from fits to $|F_\pi(s)|^2$
can be compared to data \cite{Ochs} and an effect attributed
to the $\rho(1450)$ meson seems to affect somewhat the phase
shift around $s=1$ GeV$^2$.

Beside these approaches, the most usual framework is VMD in which 
$F_\pi(s)$ is represented as a sum of vector meson contributions~;
traditionally, these are chosen as  Gounaris--Sakurai functions 
\cite{Gounaris}.
Focussing on $e^+e^-$ annihilations, this is illustrated by
the reference fit in \cite{Barkov}  to the data collected by
the OLYA,CMD and DM1 Collaborations \cite{Barkov,DM1}. The data 
set recently collected by CMD--2 \cite{CMD2} is also fitted in 
this way. In this last study, two prominent conclusions show up~:
the $\omega \rightarrow \pi \pi$ branching fraction is found smaller than
previously measured \cite{Barkov} ($1.33 \pm 0.25$ \% 
instead of $2.21\pm0.30$ \% ) and a  contribution
from the $\rho(1450)$ meson is needed in order to reach
a good description of the data set (fully located below 1 GeV).

Recently, it has been remarked  \cite{Rho0} that the Hidden Local
Symmetry (HLS) Model \cite{HLS} provides another consistent 
framework for data analysis
and a new expression for $F_\pi(s)$ at low energies. Indeed,
besides the usual vector meson exchanges, this model
predicts that some departure from standard VMD could show
up as a residual direct coupling $\gamma \pi^+ \pi^-$.
The form factor written\footnote{
We  use the so--called Orsay Phase formulation
for the isospin breaking term. This is  commented on below. }~:  

\begin{equation}
 F_\pi(s)=\displaystyle 1 -\frac{a}{2} -
 \frac{f_{\rho\gamma}g_{\rho\pi\pi}}{s-m_\rho^2 +im_\rho\Gamma_\rho(s)}
-\displaystyle \frac{f_{\omega \gamma} e^{i\phi} g_{\omega \pi \pi}}
{s-m_\omega^2+im_\omega \Gamma_\omega(s)}
\label{eq1} 
\end{equation}

\noindent has been used to fit the data then available \cite{Barkov,DM1}.
This expression provided a  nice fit \cite{Rho0} for the whole energy range 
below $s\leq 1$ GeV$^2$ without introducing any additional
vector state like the $\rho(1450)$ meson. For the HLS parameter $a$, the fit returned
$a=2.36 \pm 0.02$ in contrast with standard VMD where $a= 2$.
Here also, the phase of $F_\pi(s)$ resulting from the fit 
is a prediction for the $\delta_1^1(s)$ $\pi \pi$ phase shift
and compares well  \cite{Rho0} with the phase shift data of  \cite{Petersen}.

This model has been used, besides the usual Gounaris--Sakurai 
propagator, to fit the CMD--2 data set and it has been found to 
provide as good results \cite{CMD2}.   
In this case, the fit returned  $a=2.336 \pm 0.015 \pm 0.007$, 
in obvious correspondence with the previous estimate derived
from fit \cite{Rho0} to the former $e^+e^-$ data sets \cite{Barkov,DM1}. 
As for the previous data sets, when using the HLS model as expressed
by Eq. (\ref{eq1}), no effect below the $\phi$ mass 
was observed which could be attributed to  a  $\rho(1450)$ contribution 
in contrast with the standard (VMD) fit \cite{CMD2}. 

The aim of the present paper is to examine the pion form factor
in the context of the HLS Model, by taking into account both the
non--anomalous \cite{HLS} and anomalous \cite{FKTUY} sectors. This  
leads to consider carefully the  Analyticity requirement and to
examine the effect of the channels coupled to $\pi \pi$
within the HLS Model.  Loop effects cannot be avoided
in problems where the $\rho$ meson plays a crucial role.
These will be considered in the framework of the one--loop order
treatment proposed in \cite{mixing}. Doing this way,
one limits the possible couplings by neglecting intermediate
states with more than two particles which generate
multiparticle loops~; these are expected
to produce small effects \cite{Pich2}. This 
is supported by the experimental data of \cite{Ochs},
which exhibit a  $\pi \pi$ $P$--wave elasticity consistent 
with 1 up to about the $\phi$ mass. 

In Section \ref{Two}, we derive the pion form factor $F_\pi(s)$
in accordance with Analyticity~; we show how the $\rho$
propagator has to be dressed and that the $\gamma-\rho$ coupling
becomes invariant--mass dependent at the same order. In Section
\ref{Three}, we examine the loop corrections and show that
choosing the subtraction polynomial coefficients as fit parameters is consistent.
All this is done in the body of the text referring only to the non--anomalous
sector of the HLS Model~; more information in order to deal with the
anomalous sector are given in two Appendices.

In Section \ref{Four}, we recall the results obtained elsewhere
concerning the HLS phenomenology, which are imposed as constraints
when fitting the pion form factor. It should be noted that our
$F_\pi(s)$ has to be consistent with the $\rho$ mass derived
from the HLS--KSFR relation (827 MeV). In Section \ref{Five},
we remind which kind of information can act as (external) probes
for our HLS modelling~: the $\pi\pi$ phase shift $\delta_1^1(s)$
and the (polynomial) behaviour of $F_\pi(s)$ near $s=0$.
Our fit strategies and results on the pion form factor
are the purpose of Section \ref{Six}, while the short 
Section \ref{Seven} summarizes our fit results concerning
the $\omg$ contribution, especially Br$(\omg \ra \pi \pi)$.
Finally, Section \ref{ten} is devoted to conclusions.

\section{The Pion Form Factor In The HLS Model}
\label{Two}
 
\indent 
Actually, what comes out of the non--anomalous sector
of the HLS Model \cite{HLS} at tree level is  Eq. (\ref{eq1})
without the $\rho$ width term and amputated from the $\omega$
contribution which corresponds to some breaking of Isospin
Symmetry. Omitting these terms, Eq. (\ref{eq1}) obviously meets the
Analyticity requirement (actually, it defines a meromorphic function)
but is of little use in order to describe real data from threshold
to the $\phi$ mass. Indeed, the $\rho$ propagator which actually occurs 
there is the bare propagator $D_0(s)=(s-m_\rho^2)^{-1}$ which exhibits
a pole on the physical region $s \geq 4 m_\pi^2$.

The dressed propagator $D(s)$ is given by the Schwinger--Dyson Equation,
which writes~:

\begin{equation}
D^{-1}(s) =D_0^{-1}(s)-\Pi_{\rho \rho}(s)
\label{eq2}
\end{equation}

\noindent  at one loop order ($g^2$),
where $\Pi_{\rho \rho}$ is the $\rho$ self--energy.
Within the non--anomalous HLS Model \cite{HLS}, contributions
to the $\rho$ self--energy
come only from pion and kaon loops~; if one considers also
the anomalous sector of the HLS model, the (FKTUY) Lagrangian
of Ref. \cite{FKTUY}, additional $VP$ loops have to be introduced,
especially $\omg \pi^0$  which threshold is lower
in mass than $K \overline{K}$.

It is expected that the correct expression for the isospin
1 part of the pion form factor is obtained by replacing the
denominator in Eq. (\ref{eq1}) by the dressed propagator
$D(s)$ just defined. This can be derived by resumming formally
an obvious infinite series of terms, each containing  bare propagators
and loops (Referred to in \cite{Pich2} as Dyson--Schwinger Summation).
This expression can also be obtained by adding an effective piece 
\cite{mixing} to the HLS Lagrangian of the form $\Pi_{\rho \rho}(s) \rho^2/2$,
which turns out to modify the vector meson mass term by
a $s-$dependent piece. The (dressed) $\rho$ propagator 
is derived from this effective Lagrangian at
tree level. The Lagrangian thus defined still fulfills the hermitian 
analyticity condition \cite{ELOP}  ${\cal{L}}(s)={\cal{L}}^\dagger(s^*)$
which is the natural generalization of hermiticity.

Based on the success of analytic one--loop models \cite{Klingl,Pich2}
at energies below 1 GeV, we explore here the implications of 
extending this one--loop description to all loops permitted by 
the basic HLS model in the interest of keeping a practical 
phenomenological model.

When breaking Isospin Symmetry within the HLS Model, 
charged and neutral kaons carry different masses and this
generates a $\rho-\omg$ mass--dependent transition term\cite{su2}.
In this case, the effective piece to be added to the Lagrangian 
becomes~: 
  
\begin{equation}
\displaystyle
{\cal L}=\frac{1}{2} 
\{~\Pi_{\rho\rho}(s)~ \rho^2 + \Pi_{\omg\omg}(s) ~\omg^2
+2\Pi_{\mor}(s)~\rho \omg ~\}
\label{eq3}
\end{equation}

\noindent which implies that $\rho$ and $\omg$ mix together
and that the modified Lagrangian should be diagonalized.
It was shown in \cite{su2} that this gives rise to
an $\omg$ contribution to the pion form factor which
approximates naturally in the form shown in Eq. (\ref{eq1}), precisely.
It was also shown \cite{su2} that the proposed way of breaking Isospin
Symmetry makes the $\omg$ contribution vanishing at $s=0$ and
thus does not affect the $F_\pi(0)=1$ condition.

However, in order to stay consistent with using one--loop corrections,
the effective piece added to the Lagrangian should also contain 
loop contributions which couple the photon and vector mesons. For 
instance, the original $\rho \gamma$ term \cite{Heath1} is 
changed\footnote{We recall that the universal vector coupling $g$
is related to $g_{\rho\pi\pi}$ by $g_{\rho\pi\pi}=ag/2$ which in 
the VMD limit $a=2$ restores $g_{\rho\pi\pi}=g$. In the HLS model
an (extended) KSFR relation holds
$m_\rho^2=ag^2f_\pi^2$ and we also have $f_{\rho\gamma}=m_\rho^2/g$.} to   
$-e[f_{\rho\gamma}-\Pi_{\rho \gamma}(s)] \rho \cdot A $, where
a factor of $e$ has been extracted from $\Pi_{\rho \gamma}(s)$,
which is thus of order $g$ in couplings.

Therefore, the isospin 1 part of the pion form factor, taking into
account one--loop corrections, is~:

\begin{equation}
 F_\pi(s)=\displaystyle 1 -\frac{a}{2} -
 \frac{ \left [ f_{\rho\gamma}-\Pi_{\rho \gamma}(s)\right ] g_{\rho\pi\pi}}
 {s-m_\rho^2 -\Pi_{\rho\rho}(s)}
\label{eq4} 
\end{equation}

 In the non-anomalous sector of the HLS Model, $\Pi_{\rho \gamma}(s)$ contains 
 only pion and kaon loops as  $\Pi_{\rho\rho}(s)$. The anomalous (FKTUY) 
 part of the Lagrangian, provides additional $VP$ loops.
 We discuss in the next Section the properties of these loop corrections.
  
 The $e^+e^-$ cross
 section contains an isospin breaking term associated with the
 $\omg$ meson but also the corresponding one associated with 
 $\phi\rightarrow \pi \pi$. However, the corresponding published data 
 \cite{SND} are  not available in a usable way for fit~; fortunately,
 this effect is concentrated in a narrow region around the $\phi$ mass, 
 and is invisible in the data to be considered.  Nevetheless,
 one could note that the Orsay phase of the $\phi$ meson as well as
 its branching ratio to $\pi \pi$ are well accounted for within the
 HLS Model  broken in an appropriate way \cite{su2}. 
 
 Before closing this Section, let us remark that the $\omg$ contribution
 has practically no effect somewhat outside the $\omg$ mass region. It is
 therefore sufficient to treat it as a fixed width Breit--Wigner \cite{Rho0} with
 accepted values \cite{PDG02} for the $\omg$  mass and width and
 with a constant phase factor (see Eq. (\ref{eq1})). Additionally, we  
 neglect the effects of $\omg-\phi$ mixing by setting 
 $f_{\omg\gamma}=f_{\rho\gamma}/3
 =m_\rho^2/3g$. Taking into account the magnitude of this mixing 
 angle \cite{rad,mixing}  ($\simeq 3^\circ$ from ideal mixing), this is 
 certainly a safe assumption when fitting the pion form factor.

\section{Properties Of The One--Loop Corrections}
\label{Three}

\indent 
All loops contained in the functions   $\Pi_{\rho\rho}(s)$
and $\Pi_{\rho \gamma}(s)$ are given by Dispersion Relations
and have been computed  in closed form in \cite{mixing}. They involve $PP$
and $VP$ loops in general. Their detailed structures and the
expression of their couplings depend  on the usual HLS
parameters $g$ and $a$, but also on symmetry breaking parameters.
These have been fitted several times under various conditions
\cite{rad,mixing,su2,chpt}, always providing results consistent
with each other.

These loops should be subtracted minimally twice ($PP$)
or three times ($VP$) from requiring the corresponding 
Dispersion integrals \cite{mixing} to be convergent.
Therefore, in the full HLS Model
(non--anomalous and anomalous sectors), the subtraction polynomials
must be at least second degree in $s$ and we can write~:

\begin{equation}
\left \{
\begin{array}{lll}
\Pi_{\rho \gamma}(s)= & P_\gamma(s)+\overline{\Pi}_{\rho\gamma}(s) \\[0.3cm]
\Pi_{\rho\rho}(s) =   & P_\rho(s) + \overline{\Pi}_{\rho \rho}(s)
\end{array}
\right.
\label{eq5}
\end{equation}

\noindent where the $\overline{\Pi}(s)$ are sums of
subtracted loop functions given in \cite{mixing}, and the $P(s)$ 
are polynomials with real coefficients. 
We choose to work with second degree polynomials,
and then the coefficients to be fitted are defined by~:

\begin{equation}
\left \{
\begin{array}{lll}
P_\gamma(s)  =  & d_0+d_1s+d_2s^2  \\[0.3cm]
P_\rho(s)  =  & e_0+e_1s+e_2s^2 
\end{array}
\right.
\label{eq6}
\end{equation}

\noindent 
In this case, it is suitable to redefine the ($PP$)
$\overline{\Pi}(s)$ functions given in \cite{mixing} in such a 
way that they behave like ${\cal{O}}{(s^3)}$ near the origin.

A relevant question is whether these polynomials $P(s)$
are really independent of each other or  whether
the independent polynomials are those associated
with the pion and kaon loops contained in the  $P(s)$'s.
In this case, it is appropriate to check that
$P_\gamma(s)$ and $P_\rho(s)$ are not proportional. 

Let us discuss here only the non--anomalous sector
of the HLS model \cite{HLS}~; information given in the Appendices
allow to examine the contributions of the anomalous (FKTUY) sector 
\cite{FKTUY} with analogous conclusions. 
Using the SU(3) breaking scheme proposed in \cite{Heath1},
the piece relevant  for the pion form factor
can be extracted from Eq. (A5) in \cite{Heath1} and can be rewritten
in terms of renormalized fields ($K_{ren}=\sqrt{z} K_{bare}$,
$\pi_{ren}= \pi_{bare}$)~:

\begin{equation}
\begin{array}{lllll}
\non
{\cal L}_{\rm new}=\cdots + \displaystyle \frac{iag}{4z}\rho^0\left[
K^-\stackrel{\leftrightarrow}{\pa}K^+ -
\bar{K}^0\stackrel{\leftrightarrow}{\pa}K^0+
2z~ \pi^-\stackrel{\leftrightarrow}{\pa}\pi^+ \right] 
+
\\
 \!\!\displaystyle
{ie}A\left[\frac{(z-a/2-a(\ell_V-1)/6)}{z} K^-\stackrel{\leftrightarrow}{\pa}K^+
-\frac{a(\ell_V-1)}{6z}\bar{K}^0\stackrel{\leftrightarrow}{\pa} K^0
+(1-a/2)\pi^-\stackrel{\leftrightarrow}{\pa}\pi^+\right]
\non
\end{array}
\label{eq7}
\end{equation}

\noindent where $z$ is the SU(3) breaking parameter \footnote{
$z$ was also written $1+c_A$ in \cite{Heath1}, referring to
the original naming of \cite{BKY}, or 
$\ell_A$ in \cite{rad,mixing}.} \cite{BKY,Heath1}. 
It should be fixed to $z=[f_K/f_\pi]^2=3/2$ in order to recover
the correct value of the kaon form factor at $s=0$. Consistent
fits to radiative decay widths of light mesons confirm this value 
independently \cite{rad}. $\ell_V$ is another breaking parameter\footnote{
We have $\ell_V=(1+c_V)^2$ in terms of the original breaking parameter 
of the  ${\cal L}_V$ term of HLS Lagrangian \cite{BKY,Heath1}.}
which has also been fitted  \cite{rad} using $\omg/\phi$ leptonic decays 
as $\ell_V=1.376 \pm 0.031$. Exact SU(3) symmetry corresponds
to $z=\ell_V=1$.

Denoting $\ell_\pi(s)$ and $\ell_K(s)$ the pion and kaon loops
amputated from their couplings to external legs (we neglect
the mass difference between $K^\pm$ and $K^0$), we derive
from Lagrangian Eq. (\ref{eq7})~:

\begin{equation}
\left \{
\begin{array}{lll}
\Pi_{\rho \rho}(s)= &  \displaystyle
g^2_{\rho\pi\pi} [\ell_\pi(s) +\frac{1}{2z^2} \ell_K(s)]\\[0.3cm]
\Pi_{\rho\gamma}(s) =   & \displaystyle
g_{\rho\pi\pi} [(1-\frac{a}{2})\ell_\pi(s) +\frac{1}{2z^2}(z-\frac{a}{2})\ell_K(s)]~~~.
\end{array}
\right.
\label{eq8}
\end{equation}

Let us denote $Q_\pi(s)$ and $Q_K(s)$, the subtraction polynomials
contained in $\ell_\pi(s)$ and $\ell_K(s)$. Then, these are related
with $P_{\rho}(s)$ and $P_{\gamma}(s)$ defined above by~:

\begin{equation}
\left \{
\begin{array}{lll}
P_{\rho}(s)= &  \displaystyle
g^2_{\rho\pi\pi} [Q_\pi(s) +\frac{1}{2z^2} Q_K(s)]\\[0.3cm]
P_{\gamma}(s) =   & \displaystyle
g_{\rho\pi\pi} [(1-\frac{a}{2})Q_\pi(s) +\frac{1}{2z^2}(z-\frac{a}{2}) Q_K(s)]~~~.
\end{array}
\right.
\label{eq9}
\end{equation}

It is obvious that the single case where $P_{\rho}(s)$ and 
$P_{\gamma}(s)$ are not independent of each other
is if $z=1$ (no breaking of SU(3) symmetry)~; then  
$e_i = g_{\rho\pi\pi} d_i$. 
Therefore, it is quite legitimate to treat
$P_{\rho}(s)$ and $P_{\gamma}(s)$ as independent when
fitting experimental data.

It is usual and motivated to assume that the constant terms
of the polynomials $Q_\pi(s)$ and $Q_K(s)$ are zero in
order to ensure masslessness of the photon, after dressing
its propagator (see Ref. \cite{mixing} and references
quoted herein)~; this turns out to fix $ d_0=e_0=0$
in Eqs. (\ref{eq6}).
We shall keep this assumption throughout this paper
\footnote{Assuming  non--zero $ d_0$ and $e_0$, would be  
{\it practically} equivalent to releasing any constraint on
$m_\rho$ and $f_{\rho \gamma}$ as clear from Eq. (\ref{eq4}).
 }.

\vspace{0.5cm}

\indent
Some additional remarks are of relevance before closing this Section.
Within standard VMD ($a=2$), the $\rho$ propagator is still 
dressed by loop effects as described above. However,
we could also expect that no one--loop dressing connects
the intermediate photon with the $\rho$ meson and therefore
$\Pi_{\rho \gamma}$ would disappear from the form factor Eq. (\ref{eq4}).
The equations just above show that this statement is not true,
as~:
\begin{equation}
\lim_{a\rightarrow 2} \Pi_{\rho \gamma}(s)=\frac{g}{2 z^2}(z-1)\ell_K(s)
\label{eq10}
\end{equation}

Therefore, an invariant--mass dependent  dressing of the $\rho-\gamma$ 
coupling occurs as consequence of SU(3) symmetry breaking 
of the HLS model and this statement is valid 
for all proposed breaking schemes\cite{BKY,Heath1,BGP} 
of the HLS Model\footnote{It is interesting to note
that the phase of $F_\pi(s)$ in Eq.(\ref{eq4}) is
given by only the denominator, up to the first inelastic threshold.
In the non--anomalous HLS Model this is $K\overline{K}$
and then an imaginary part is generated by a term
identical to the one written down in  Eq.(\ref{eq10}) above the 
$\phi$ mass. If one adds the anomalous sector \cite{FKTUY}, things
are somewhat different, as the lowest inelastic threshold becomes
 $\omg \pi^0$.}.  Additionally, it implies that
assuming  VMD ($a=2$), the HLS model looses
its direct $\gamma \pi^+ \pi^-$ coupling, but 
SU(3) breaking generates direct $\gamma K^+ K^-$
and $\gamma K^0 \overline{K}^0$ couplings.
 
A specific character of the HLS model is that it contains
a direct coupling of the photon to pseudoscalar pairs and this generates
a mass--dependent dressing of the $\gamma -\rho$ transition.
However, this property is shared with another identified class
of models named VMD1 in \cite{Rho0} (for a review, see \cite{vdm}).  
A first such model which illustrates
that loop dressing of the $\gamma -\rho$ transition can accomodate
pion  form factor data is given in \cite{Klingl}~; quite recently,
the same idea was developed up to a more refined comparison
with experimental data up to the $\phi$ mass \cite{Melikhov}.
We note that it has been explicitly demonstrated that
regular VMD and VMD1 are equivalent \cite{Tony}, as one would expect
from corresponding fit results \cite{Rho0}.

\section{Phenomenological Constraints On $F_\pi(s)$}
\label{Four}

\indent The HLS model \cite{HLS} depends basically on only two parameters
to be determined experimentally. The universal vector coupling constant
$g$ and the parameter $a$ which allows to extend the model beyond the standard
VMD assumption corresponding to $a=2$. In principle, its Lagrangian
gives predictions  for all hadronic two--body decays of light pseudoscalar 
and vector mesons. However, in order to describe experimental data,
schemes providing the HLS Model with symmetry breakings are unavoidable.

Only a few physics processes can be phenomenologically accounted for
without significant symmetry breaking effects, noticeably the pion form 
factor. Simply using a varying width Breit--Wigner formula\footnote{
This does not fulfill the requirement of Analyticity.}
 for the
$\rho$ propagator, the HLS Model can achieve a quite satisfactory description
of $F_\pi(s)$ from threshold to the $\phi$ mass \cite{Rho0,CMD2}. This
description compares well with other approaches accounting for the 
Analyticity requirement \cite{Yndurain,Pich1,Pich2,Klingl,Gounaris,Melikhov} 
or not \cite{Rho0}. Actually, from a phenomenological point of view, the 
Analyticity assumption for $F_\pi(s)$ gets its full importance only when 
predictions outside have to be derived from timelike region data and 
fits~: in the spacelike region or  near the  chiral point. It was
indeed shown in  \cite{Rho0} that the behaviour of $F_\pi(s)$ near
$s = 4 m_\pi^2$ predicted from ChPT was well accounted for and that
its phase describes quite well the $\delta_1^1(s)$ phase shift up
to the $\phi$ mass. 

Therefore, even if successfull with  $F_\pi(s)$, establishing firmly the HLS Model
as a consistent framework for physics analysis needs further confirmation.
To extend the range of experimental data accessible to the HLS model,
a consistent  SU(3) Symmetry breaking scheme was provided \cite{BKY,Heath1}
and also a scheme for breaking of Nonet Symmetry \cite{rad}. Supplementing
the non--anomalous HLS Lagrangian \cite{HLS} with its anomalous sector 
\cite{FKTUY}, it was then possible to prove that all radiative and
leptonic decays of light mesons were successfully described within the
HLS framework~; additionally, it was shown \cite{chpt} that this 
framework meets all expectations of Extended ChPT \cite{Kaiser}
concerning decay constants and the mixing angle $\theta_8$.
The value derived from our fits for $ \theta_0=-0.05^\circ \pm 0.99^\circ$
did not match well with the leading order ChPT estimate \cite{Kaiser}
$ \theta_0 \simeq -4^\circ$, however, a recent next--to--leading
order calculation \cite{Holstein}  ($\theta_0=[-2.5^\circ,+0.5^\circ]$)
restores agreement  with its phenomenologically extracted value.

For thorough discussions on the phenomenological results
derived from  the broken HLS Model, we refer the reader to 
\cite{rad,mixing,chpt,su2}. Specific information concerning
the pion form factor are given in the Appendices.
Here we focus on discussing the parameters entering
explicitly Eq. (\ref{eq4}) and the coupling constants affecting
the non--anomalous Lagrangian Eq. (\ref{eq7})~: $a$, $g$ 
and $z$~; in the limit of unbroken Isospin Symmetry,
the breaking parameter $\ell_V$ drops out from the 
pion form factor expression.

Pion form factor fits \cite{Rho0,CMD2} give two measurements
consistent with each other which can be averaged 
as $a=2.35 \pm 0.01$. From a global fit of all radiative and 
leptonic decays of light meson \cite{rad}, the best fit value is 
$a=2.51\pm 0.03$. Variants of this model with a mass dependent 
$\omg-\phi$ mixing angle \cite{mixing}, or  accounting for 
isospin breaking effects \cite{su2} give values consistent with 
this one at never more than $2 ~ \sigma$.

There is a significant departure between the value of 
$a$ derived from fits to the pion form factor and the
value coming from fit to radiative and leptonic decays. As noted in
\cite{CMD2}, below $s=1$ GeV$^2$, it could be hard
to disentangle completely effects of departures from
strict VMD ($a=2$) and effects of resonance tails
(namely, the $\rho(1450)$). The global fit to radiative
and leptonic decays can be considered as safer from this point 
of view and then it looks well founded to prefer using 
$a=2.51\pm 0.03$.
This turns out to attribute the difference with $a=2.35$
to higher resonance effects not accounted for in the HLS
fits in \cite{Rho0,CMD2} and/or to another phenomenon
(mass dependent $\rho-\gamma$ coupling).

All fits to the data considered \cite{rad,mixing,chpt,su2} 
return $ g=5.65 \pm 0.02 $. Finally, fitting
the SU(3) breaking parameter $z$ within this data 
set \cite{rad,su2} always returned $z=[f_K/f_\pi]^2=3/2$,
as also expected from $F_K(0)=1$ \cite{BKY,Heath1}.

If a consistent picture of the HLS phenomenology
can be achieved, it implies that these parameters
can be fixed at the values corresponding to the
best fit in radiative and leptonic decays (values 
recalled just above).
In this case, the only parameters relevant for the $\rho$ meson 
which can be allowed to vary are the (non--identically zero) 
coefficients of the subtraction polynomials in Eq. (\ref{eq6}).
Indeed, the HLS Model satisfies a modified KSFR
relation which fixes the $\rho$ mass, $m_\rho^2=a g^2 f_\pi^2$
and $f_{\rho\gamma}=m_\rho^2/g$ in terms of only $a$ and $g$.
As we have neglected the $\omg -\phi$ mixing mechanism,
$\omg$ is approximated by its ideal component and then
$f_{\omg\gamma}=f_{\rho\gamma}/3$ is also fixed.
 
Therefore, it is
a kind of global fit to radiative and
leptonic meson decays and to the pion form factor  
to fit $F_\pi(s)$ by fixing $a$, $g$ and $z$.
However, this means that the $\rho$ mass is fixed
to $m_\rho=827 \pm 4$ MeV~; using the relation 
between $g_{\rho \pi \pi}$ and the width, the 
$\rho$ width would correspond to 
$\Gamma_\rho \simeq 135 $ MeV. 

Both values are clearly far from matching expectations 
\cite{PDG02} and one may wonder how the pion form
factor could accomodate such $\rho$ parameters.
However, as noted in \cite{mixing}, finite
width effects ({\it i.e}, loop corrections) should
perform the consistency. One aim of the present
paper is to check and show that all consequences on $\rho$ 
parameters of the radiative and leptonic decays are indeed 
accomodated by the pion form factor. 

It is also important to point out a couple of subtleties.
The $\rho$ mass, as defined by the real part of the propagator 
$M_\rho$, is highly model dependent \cite{Gardner}. The complex pole, however, 
is a true invariant, as has been shown for several models \cite{Tony}.
One should also note the difference between $M_\rho$ and the 
``Higgs--Kibble" mass $m_{HK}$ ($m_\rho^2=m_{HK}^2=ag^2f^2_\pi$)
\cite{Tony} resulting from spontaneous symmetry breaking.

\section{Probes and Data Sets}
\label{Five}

\indent
Any fit performed with Eq. (\ref{eq4}) actually
returns an analytic function with some uncertainty
on the fit parameters. These fits always optimize
the description of data sensitive to only $|F_\pi(s)|$.

A first probe, as for other studies (see \cite{Yndurain}
for instance), is to compare the phase {\it predicted} by
$\rm{Arg}[F_\pi(s)]$ with the most reliable data
on the $\delta_1^1(s)$ phase shift \cite{Ochs,Protopescu}
below $\simeq \sqrt{s} \simeq 1 $ GeV.

A second probe is to compare numerically the behaviour 
of this fitted $F_\pi(s)$ near $s=0$ to external sources. These are
mostly  ChPT predictions \cite{GL,Bijnens} or approches relying
on the inverse amplitude methods \cite{Hannah,Truong,Yndurain,Pich1}.

Defining the low energy expansion of $F_\pi(s)$ by~: 

\begin{equation}
F_\pi(s)= 1 + \lambda_1 s + \lambda_2s^2+\lambda_3 s^3 + \cdots
\label{eq11}
\end{equation}

\noindent the works just quoted find parameter values as given in Table \ref{T1}~;
the results of \cite{Hannah} are very close to those  displayed and do not quote
estimated errors. We also display the results of polynomial fits \cite{Davier} 
to timelike data ($\sqrt{s} \leq 0.6$ GeV), fixing the charge radius 
($<r_\pi^2>=6 \lambda_1$) to the value found by the NA7 Collaboration \cite{NA7}.

 It is clear from  Table \ref{T1} that, whatever the method, there is an overall
 consensus about  the value of $\lambda_1$. Even if not as nice, the agreement 
 for the value of $ \lambda_2$ looks quite reasonable. The spread of central 
 values for $\lambda_3$  and their accuracies should be however noted. It 
 indicates, at least, some model dependence.
 
 \vspace{0.5cm}
 
 \indent
 The data sets which basically enter our fits are the former \cite{Barkov}
 and recent \cite{CMD2} data on $e^+ e^- \ra \pi^+ \pi^-$ together and
 separately. For convenience, the $\tau$ data \cite{Aleph,Opal,Cleo} are not
 considered in the present paper.
 Additionally, we limit our fits to the region
 below $s=1$ GeV$^2$, for reasons to be explained just below.
 
 We also consider the spacelike form factor data of NA7 \cite{NA7} and 
 of the Fermilab experiment 
 \cite{fermilab2} after some check of (fit) consistency with
 the timelike data. With these data, our fit region extends from
 $s \simeq -0.25$ to $s \simeq 1$ GeV$^2$.
 
 Finally, we will compare the phase of $F_\pi(s)$ {\it derived} from 
 fitting $|F_\pi(s)|$ to the phase shift data of \cite{Ochs,Protopescu}.
 These last data sets will be used as probes and not included in the fitted data.
 
 \indent
 For the time being, we also do not attempt to extend our fit
 (and/or the HLS Model) to higher $s$ values (namely, above the $\phi$ mass), 
 where effects of $\rho(1450)$ and $\rho(1700)$ have certainly
 to be accounted for. Extending the HLS Model to such energies
 is an interesting issue, however, it is not clear whether we would 
 not be going beyond the validity range of the HLS Model which is a low
 energy model.

\section{Fitting The Pion Form Factor}
\label{Six}

\indent In several preliminary studies, we tried examining
the behaviour of our fit parameters to the former
\cite{Barkov} and recent \cite{CMD2} $e^+e^-$ data.
All fit parameters have been found quite insensitive
to any difference, except for the $\omg$ branching
fraction to $\pi^+\pi^-$ and the Orsay phase~; this   
particular point will be examined in Section \ref{Seven}.
Therefore, in this whole Section, we consider
together the data collected in  \cite{Barkov} 
and \cite{CMD2}.

The effect of considering the timelike data
\cite{Barkov,CMD2} in isolation and combined
with spacelike data \cite{NA7,fermilab2} is
more noticeable and amounts to about a $2 \sigma$
deviation.  Nevertheless, this effect is limited
and these data sets contribute to improving the behaviour
of the pion form factor in the spacelike region
by avoiding to extrapolate without any information.
Therefore, for the work reported in this Section,
we have prefered keeping them in the 
data set to fit, reestimating the errors  
correspondingly.

\subsection{Fit Strategies and Properties}

\indent 
We report in the following on various strategies 
used to fit the pion form factor. These  differ only by
a progressive account of all permitted coupled channels.
We stress once again that the number of fitted data points
is always the same and that the number of free parameters
in the fit is not modified when accounting for more and more coupled channels.
We always have the 4 non--zero subtraction parameters 
defined in Eq. (\ref{eq6}) which account for the $\rho$ contribution
and 2 more parameters to account for the $\omg$ contribution (named
$\phi$ and $g_{\omg \pi \pi}$ in Eq. (\ref{eq1})).

Qualitatively, all fits give always large  
correlation (above the 95\% level) 
coefficients $(e_1,e_2)$ and $(d_1,d_2)$. However, the
correlation coefficients between both sets is always in the range 
of 10 to 40 \%. The parameters defined in Eq. (\ref{eq11})
are derived by expanding Eq. (\ref{eq4}) near $s=0$~;
when computing errors on the $\lambda_i$, 
errors and correlations on fitting parameters are taken 
into account. 

On the other hand, the fit qualities associated with
subsets of possible coupled channels are displayed as last
data column in Table \ref{T2}. They clearly show, that the fit
quality reached below the $\phi$ mass is always good.
From examining the evolution of the minimum  $\chi^2$, 
one should note that adding $K \overline{K}$ gives
no improvement or no degradation in the model description.
In contrast, one could remark the jump in  probability  when adding 
the $\omg \pi^0$ channel~; indeed, it is a noticeable effect 
to reduce the minium $\chi^2$  from $\simeq 184$ to $174$ without any 
additional freedom in the model. This clearly comes from a better 
account of the pion form factor between the $\omg \pi^0$ threshold
and the $\phi$ mass where data are affected by small errors.
Instead, the $K \overline{K}$  channels can noticeably
affect only a very few data points~; their very small 
effect might be due to the fact that the corresponding loops 
are numerically negligible when computed very close to threshold.

One could also note that adding the higher $VP$ coupled channels 
goes on slightly improving the fit quality below the $\phi$ mass~; 
as stressed above, this does not correspond to having more freedom 
in the model. In contrast with the $\omg \pi^0$ channel, effects
of these higher threshold loops are modest and can be 
neglected\footnote{Whether adding $K^* K$ channels and higher threshold 
channels is appropriate, while neglecting high mass meson contributions 
or multiparticle loops, can certainly be questioned.
}. One might note, however, that these have more effects on data 
than the $K \overline{K}$ channels, as clear from Table \ref{T2}.

Finally, we have attempted fits by removing
the function $\Pi_{\rho \gamma} $ from Eq. (\ref{eq4}),
while keeping all parameters fixed by HLS phenomenology 
at their values obtained from fit to radiative 
and leptonic decays ($a$, $g$, \ldots).
We never reached a reasonable result. In order to remove
this function one clearly needs  to release (at least a part 
of) these constraints in order to allow the fit to converge
to a good description of the data{\footnote{
The fit quality and results in \cite{Pich2,Klingl,Melikhov} 
clearly proves this statement.}}.
This was not attempted as our aim was to examine the full
consequences of having the HLS Model {\it and} all known numerical
constraints coming from its phenomenology. We thus perform
a kind of global fit of all relevant decay modes
and of the pion form factor together. This exercise, however,
tends to indicate that the dressing of the $\rho-\gamma$
transition amplitude is a relevant concept and is requested
by the consistency of the pion form factor with the rest
of the HLS phenomenology.

\subsection{The Pion Form Factor in Spacelike and Timelike Regions}

\indent  As stated above, whatever the subset of channels considered,
the last data column in Table \ref{T2} shows that the fit quality is 
optimum in both the spacelike and timelike $s$ regions considered.
Accounting for the three coupled channels $\pi^+ \pi^-$, $\omg \pi^0$, 
$K \overline{K}$ (actually the neutral and charged modes) seems
the best motivated coupling scheme for the invariant--mass region
from threshold up to the $\phi$ mass. We thus illustrate with
this our fit results~; visual differences with other channel 
subsets are tiny. 

The fit functions discussed 
in all this Subsection have been obtained from a global fit to
all existing timelike data \cite{Barkov,DM1,CMD2} and
to the spacelike data of \cite{NA7,fermilab2} simultaneously.

\begin{figure}[htb]
  \centering{\
     \epsfig{angle=0,figure=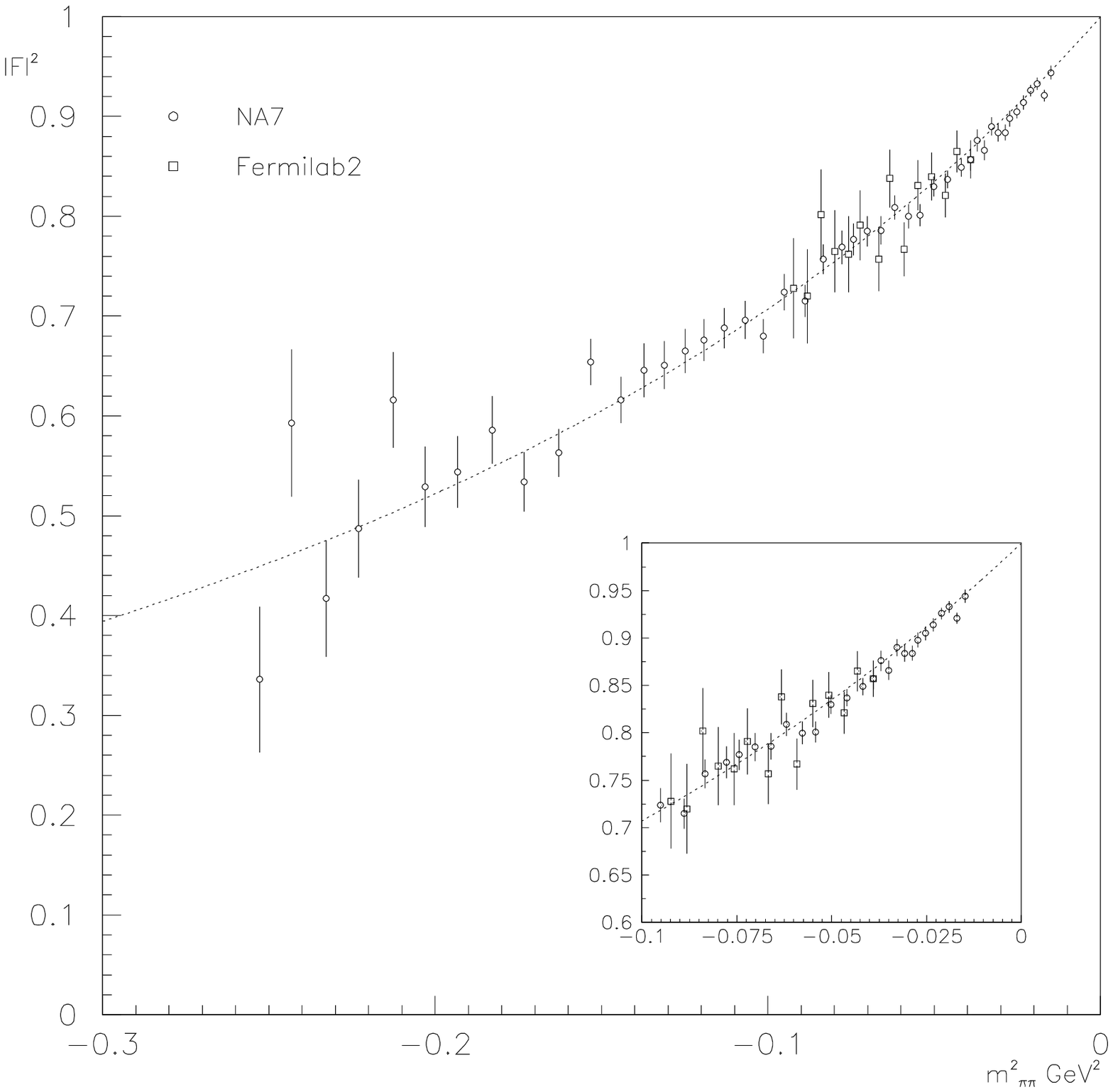,width=0.9\linewidth}
                    }
\parbox{150mm}{\caption ~~~
Spacelike data and fit. The data points are from Refs.
\cite{NA7} and \cite{fermilab2}. The fitting curve
has been obtained by considering the $\pi \pi$,
$K \overline{K}$ and $\omg \pi^0$ channels. All
channels subsets as defined in the body of the text
(including $\pi \pi$ alone) give representations
hard to distinguish from the one shown.
\label{spacelike}
}
\end{figure}

In Fig. \ref{spacelike}, we display the fitted form factor
in the spacelike region and, superimposed, are the data of 
\cite{NA7,fermilab2}. As expected, the description is quite 
reasonable.

\begin{figure}[htb]
  \centering{\
     \epsfig{angle=0,figure=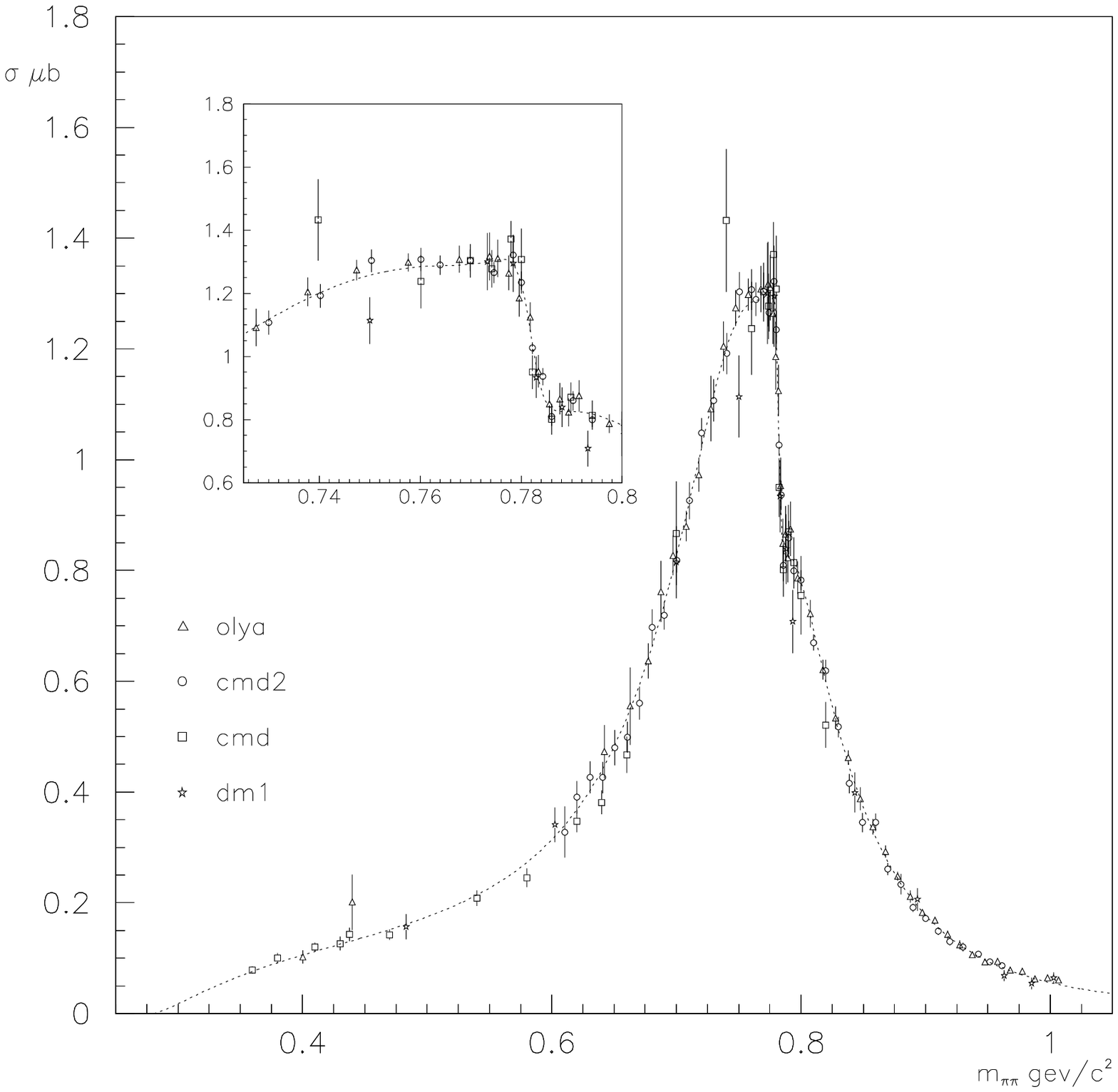,width=0.9\linewidth}
                    }
\parbox{150mm}{\caption ~~~
Timelike data and fit. The data points are 
all subsets from Refs.
\cite{Barkov,DM1,CMD2}. The fitting curve
has been obtained by considering the $\pi \pi$,
$K \overline{K}$ and $\omg \pi^0$ channels. All
channels subsets as defined in the body of the text
(including $\pi \pi$ alone) give representations
visually identical to the one shown here.
\label{timelike}
}
\end{figure}

\begin{figure}[htb]
  \centering{\
     \epsfig{angle=0,figure=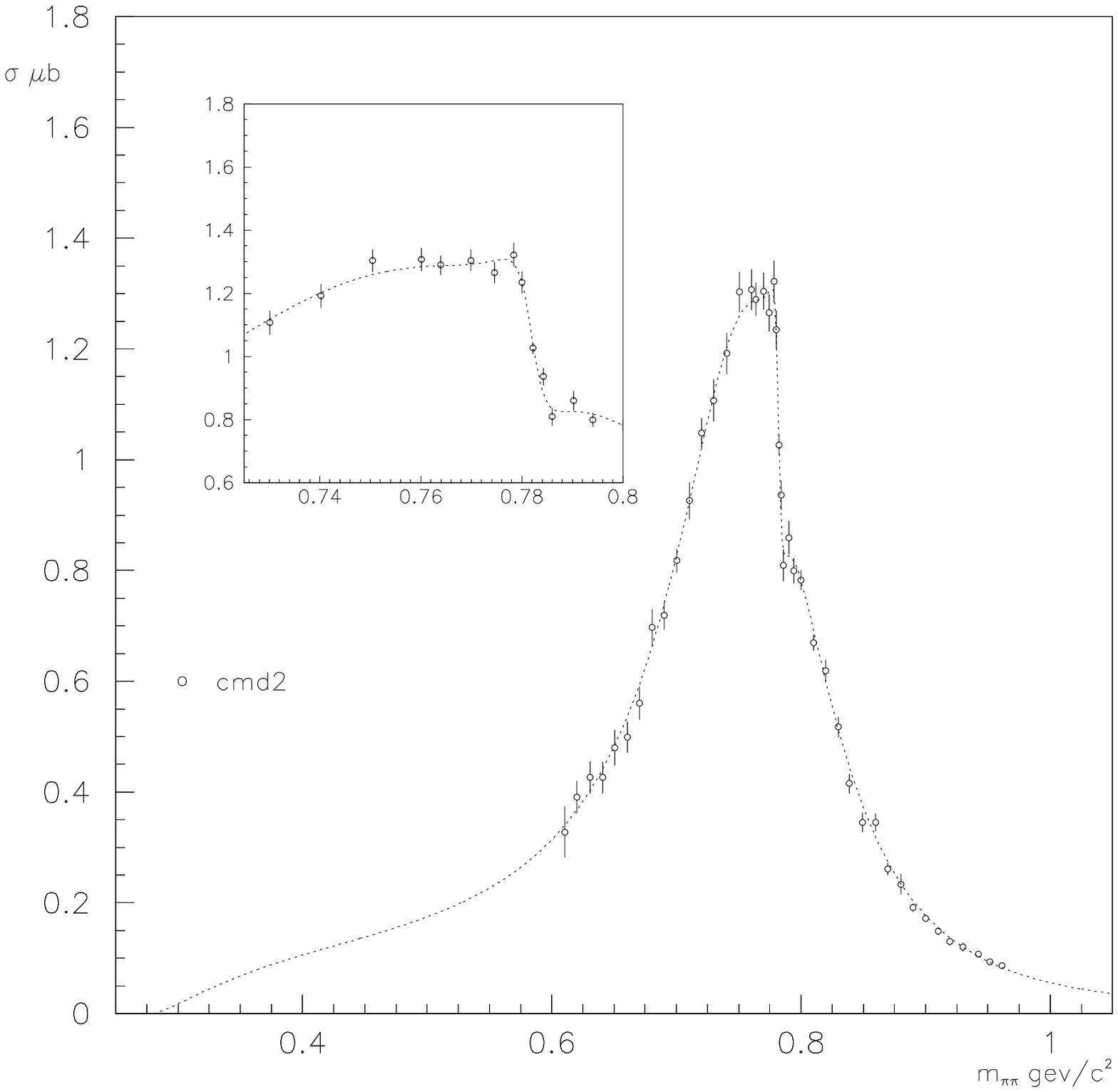,width=0.9\linewidth}
                    }
\parbox{150mm}{\caption ~~~
Timelike data and fit. The data points are only
from the recent  data set collected by the CMD--2 Collaboration
\cite{CMD2}. The fitting curve is the same as in Fig. 
\ref{timelike} and its numerical coefficients have been 
determined by a global fit to all available timelike data
and to the spacelike data of \cite{NA7,fermilab2}.
\label{timelikenew}
}
\end{figure}

In Fig. \ref{timelike}, we show the fit in the timelike region 
superimposed with all existing data \cite{Barkov,DM1,CMD2}~;
in Fig \ref{timelikenew}, we have displayed the same fitting
curve with only the new data of \cite{CMD2}.
Fig.  \ref{timelike} shows that the whole energy range is well described, 
including the region below 600 MeV (measured by CMD). 
Fig. \ref{timelikenew} illustrates that the fitting curve derived
from fitting all timelike data altogether give also  a very
good description of the recent CMD--2 data set \cite{CMD2} alone in its full range. 
It should be noted that, in this case, the fitting function corresponds to 
Br$(\omg \ra \pi^+ \pi^-)= 2.12 \pm 0.23$ \%~; we comment more on this 
point in Section \ref{Seven}.

It should also be remarked that the fitting function being an analytic
function of $s$, it is the same function (given in Eq. (\ref{eq4}) and
supplemented with the last term from Eq. (\ref{eq1}) to account
for the $\omega$ mass region) which fits the spacelike and timelike
$s$ regions simultaneously. 

\vspace{0.5cm}

The noticeable peculiarity of CMD--2 data with respect to the previous
data sets is that their systematic errors are smaller than 1\% \cite{CMD2};
from what is shown here, one could conclude that 
the previous data sets, considered altogether, behave globally with
small effective systematics. It could also be that the fitting function
is analytically enough constrained to be marginally sensitive
to systematics. 

{From} what is just discussed, we already know that the data description
following from our model is quite reasonable. As clear from Figs.
\ref{timelike} and \ref{timelikenew}, no need for a $\rho(1450)$ 
contribution shows up below 1 GeV. This is also illustrated
by the fit quality already reached in all cases (see the
last data column in Table \ref{T2}).

As a final remark, one should note that the high value for $m_\rho=827$
MeV is not inconsistent with the data, provided the model pion form factor is
suitably parametrized.  We  noted already that this mass value derived from HLS 
phenomenology \cite{rad} is very close to the estimate of the vector nonet 
mass fitted in \cite{Pich2}. This proves, as noted in \cite{mixing},
that it is the loop effects which are responsible of pushing the peak 
location of the $\rho$ meson (or its pole location) to the customary 
value attributed to its mass \cite{PDG02}.

Now, we focus on comparing
refined consequences of our model and fits with numerical predictions
concerning the behaviour of the pion form factor at threshold,
and the phase of our fitted $F_\pi(s)$.

\subsection{Pion Form Factor Behaviour At Threshold}

\indent 
The results we got concerning the pion form
factor behaviour at threshold are gathered in Table \ref{T2}.
They are displayed using the notations of Eq. (\ref{eq11}).
Each line corresponds to a case where a subset of the coupled
channels is considered and the size of the coupled channel subset
is increased. The second line breaks the obvious rule but is given in 
order to show that coupling the $K \overline{K}$ has negligible 
numerical effects on $F_\pi(s)$ and does not improve or degrade 
the description obtained using only the $\pi \pi$ channel.

A first remark which can be drawn is that $\lambda_1$
(hence, the pion charge radius) is totally insensitive
to whatever we add to the $\pi \pi$ channel. For this parameter,
our estimate~: 

\begin{equation}
\lambda_1= 1.896 \pm [0.018]_{stat} \pm [0.03]_{syst.}~~ {\rm GeV}^{-2}
\label{c1}
\end{equation}
 
\noindent is in good agreement with all reported values~:
 ChPT at two--loops result \cite{Bijnens}, phase
method result of \cite{Yndurain}, resonance ChPT result \cite{Pich2},
or the inverse amplitude result of \cite{Truong} (see Table \ref{T1}). 
The systematic error is estimated by considering the variation
of the central value of $\lambda_1$ as a function of the subset
of coupled channels.

The second coefficient $\lambda_2$ varies only little
as function of the number of open channels. However,
there is a clear systematic effect~: its value decreases
slowly when new channels are opened (except for  
$K \overline{K}$ commented on above). Interestingly,
the values we get always match nicely several
entries in Table   \ref{T1}. This column in
Table \ref{T2} leads us to conclude~:

\begin{equation}
\lambda_2= 3.85 \pm [0.06]_{stat.} \pm [0.10]_{syst.}  ~~ {\rm GeV}^{-4}
\label{c2}
\end{equation}

\noindent where the systematic error is estimated as for $\lambda_1$.
This result matches well expectations from Table \ref{T1}.

For the third coefficient $\lambda_3$, the situation is much
more embarassing\footnote{From Table \ref{T1} alone, the situation
looks already confusing, even by leaving aside the result of Ref.
\cite{Davier}.}. One should note that $\lambda_3$ depends on
the fit parameters ($e_i$ and $d_i$), and also on the third
order term of the loops. This third order term is fixed and given by the driving 
terms of all loops. The only way to change it is to oversubtract 
the loops and introduce (free) $e_3 s^3$ and $d_3 s^3$ terms in Eqs. (\ref{eq6})
to be fitted and/or fixed. However, the fit quality already reached with fitting
data below $\sqrt{s}=1$ GeV cannot justify to simply increase the 
model freedom. It thus seems that a reliable estimate of $\lambda_3$ 
depends on a reliable account of data somewhat above the $\phi$ mass
and on other sources of inelasticity  generally neglected,
like multiparticle loops.  

Anyway, one should note first that the first two terms of the chiral
expansion of $F_\pi(s)$ are well defined and this is not changed (or spoiled)
by adding more and more coupled channels.  Secondly, one can assess that
the data bounded to $\sqrt{s} \leq 1$ GeV alone look unable
to permit a real measurement of $\lambda_3$, as its central value
sharply depends on the inelasticity accounted for in the
region  $\sqrt{s} \geq 1$ GeV. This inelasticity was here
represented by high mass channels coupling to the $\rho(770)$ meson,
however it could have been anything else (like higher $\rho$ meson 
contributions). Stated otherwise, without a reasonably good knowledge 
of (generally neglected) inelasticity effects, the pion form
cannot provide a reliable estimate of  $\lambda_3$.

\subsection{The phase of $F_\pi(s)$ and Phase shift Data}

\indent
As stated above several times, all numerical
parameters of the analytic function $F_\pi(s)$ --
actually, only its isospin 1 part is relevant here-- are derived
from fits to data sensitive only to $|F_\pi(s)|$.
Therefore Arg$[F_\pi^{I=1}(s)]$ is a prediction and can be compared
with the most precise experimental information on the phase
shift $\delta_1^1(s)$ \cite{Ochs,Protopescu}.

\begin{figure}[htb]
  \centering{\
      \epsfig{angle=0,figure=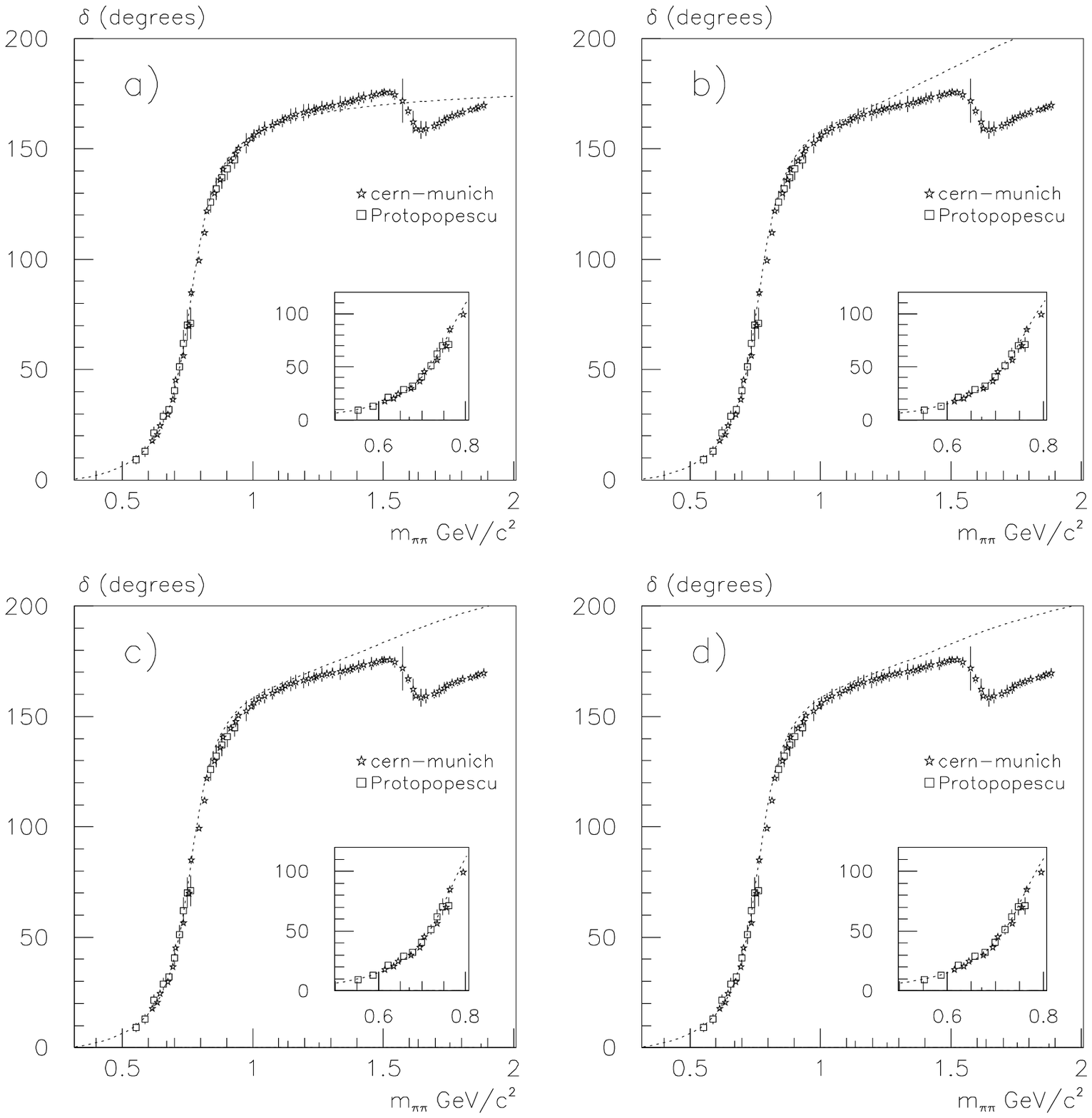,width=\linewidth}
                    }
\parbox{150mm}{\caption ~~~
Comparison with the $\pi \pi$ phase shift data of
\cite{Ochs} and \cite{Protopescu}. The curve plotted
is Arg$[F_\pi(s)]$ with parameters fixed at values
corresponding to the best fit of $|F_\pi(s)|$ 
using all timelike data \cite{Barkov,DM1,CMD2} and
the spacelike data from \cite{NA7,fermilab2}.
In {\bf a}, only the $\pi^+ \pi^-$ channel is considered~;
in {\bf b}, the subset considered is $\pi^+ \pi^-$,
$\omg \pi^0$ and both $K \overline{K}$ channels. In 
{\bf c}, the four  $K^* K$ channels have been added 
to the previous channel subset~; in {\bf d}, the
previous subset is extended so as to include 
$\rho^0 \eta$ and $\rho^0 \eta^\prime$. The agreement
is perfect up to $\simeq 800$ MeV and good up to
$\simeq 1.2$ GeV always.
\label{phase}
}
\end{figure}

In Fig. \ref{phase}, we display this comparison using
coupling to only $\pi \pi$ (Fig. (\ref{phase} a)), then
coupling to both $\pi \pi$ and $\omg \pi$ (Fig. (\ref{phase} b)~;
these do not differ from their partners with also the $K \overline{K}$
channels opened. In Fig. (\ref{phase} c), the open channels
are all channels up to the 4 contributing $K^* K$ final states; finally,
in Fig. (\ref{phase} d), all possible channels of the full 
HLS model are considered (the previous subset plus $\rho \eta$
and $\rho \eta^\prime$). 

In all cases, the insets show that the low energy region
is perfectly predicted up to $m_{\pi \pi} \simeq 800$ MeV,
whatever the subset of coupled channels considered.

Using coupling to only $\pi \pi$ (Fig. (\ref{phase} a)), the
agreement between our prediction and data is perfect up to
about 800 MeV and remains very good up to $\sqrt{s} \simeq 1.3$ GeV. 
Adding  $K \overline{K}$ does not modify sensitively this picture.

As soon as one opens the $\omg \pi$ channel, the predicted phase 
starts to diverge almost linearly from the experimental data of \cite{Ochs} 
from about  $m_{\pi \pi} \simeq 1.2$ GeV.
Nevertheless, the phase remains perfectly reproduced up to 
$m_{\pi \pi} \simeq 0.8$ GeV. From about 900 MeV, the predicted
phase starts running 2 to 4 degrees above the data of \cite{Ochs}~; this effect
is systematic but consistent with the data.
It is worth remarking that the  first inelastic  coupled channel
in the full HLS model is $\omg \pi^0$ with  threshold
 located at 917 MeV. Therefore, from Watson theorem, one
can indeed expect that  the phase predicted by the pion form factor
and the $\delta_1^1(s)$ phase shift should start 
diverging\footnote{Actually, it is not that much the divergence between
the  phase of  $F_\pi(s)$ and the $\delta_1^1(s)$ phase shift 
above 917 MeV which looks appealing. It is rather the agreement
between them up to  $m_{\pi \pi} \simeq 1.3$ GeV when limiting
the subset of coupled channels to $\pi  \pi$ and $K \overline{K}$
which could look unphysical. However, examining the elasticity
of this wave \cite{Ochs} indicates that the $(I=1,~~l=1)$  $\pi \pi$ 
wave is still elastic at a $\simeq $ 95 \% level at this energy.
Therefore, nothing conclusive can be derived from this  unexpected
agreement at relatively large invariant--mass.} at $m_{\pi \pi}=917$ MeV.

Keeping in mind the words of caution already stated concerning the 
appropriateness of considering too high threshold mass channels,
it is nevertheless interesting to remark a curious effect
of the corresponding inelasticity~: the quasi--linear rise of the phase
above 1.2 GeV which follows from having introduced the coupling
to $\omg \pi^0$  is  softened more and more, when more
(high mass) coupled channels are considered.

We remarked already that our fits to annihilation data below the 
$\phi$ mass do not exhibit any failure which could be attributed to
some neglected $\rho(1450)$ contribution. On the other hand,
because we have no guide like the Watson theorem, nothing
clear can be stated by observing the higher energy behaviour 
of the predicted phase when accounting for $VP$ loops.
However, the continuation of the annihilation cross section above the 
$\phi$ mass becomes too large when  $VP$ loops are accounted for.
Therefore, if fitting some mass region above the $\phi$ meson,
beside introducing the $\rho(1450)$ and 
$\rho(1700)$ mesons, one certainly needs to modify
the subtraction scheme by going to higher degree
subtraction polynomials. This issue will not be examined
any further here.

\section{The $\omg$ Information from Fits}
\label{Seven}

\indent  We stated in the previous Section that there
was no noticeable difference between the former annihilation 
data sets considered together and the new data set, as far as
the isospin 1 part of the pion form factor is concerned.
However, this does not extend to the $\omg$ parameters 
accessible from the pion form factor in the timelike region.

We have performed several fits and we report
on using the set of all open channels. By closing
the high energy ones, one does not change the picture 
described now.

A fit to all former $\pi \pi$ timelike data \cite{Barkov,DM1}
gives~:

\begin{equation}
\begin{array}{llll}
{\rm Br}(\omg \ra \pi \pi) = 2.27 \pm 0.35 {\rm \%} ~~~, & \varphi = 106.87^\circ \pm 7.16^\circ
\end{array}
\label{old}
\end{equation}

\noindent with $\chi^2/dof= 63.63/76=0.84$ corresponding to a 84\% probability.
It is close to the accepted value of $2.21\pm 0.30$ \%.
On the other hand, the same fit to the new data set of CMD--2 \cite{CMD2}
provides~:

\begin{equation}
\begin{array}{llll}
{\rm Br}(\omg \ra \pi \pi) = 2.01 \pm 0.29 {\rm \%} ~~~, & \varphi = 103.88^\circ \pm 2.91^\circ
\end{array}
\label{new}
\end{equation}

\noindent with $\chi^2/dof=32.20/37=0.87$ corresponding to a 69\%  probability.
This has to be compared with the result recently published
by CMD--2 Collaboration \cite{CMD2} which finds $1.33 \pm 0.25$ \% from their
fits. The central values of this result and ours are far apart 
(however, a $2 \sigma$ deviation only)~; this might also
illustrate some model dependence in extracting this information\footnote{
It should be remarked that our fit of the data collected in \cite{Barkov}
gives a result close to the published fit of OLAY and CMD data (namely
$2.30 \pm 0.5$ \%). For this fit, \cite{Barkov} was taking into account 
the coupling to $\omg \pi^0$ channel in the way proposed by \cite{costa}.
Fitting the former data in \cite{Barkov} as done now with the new data
gives instead $2.00 \pm 0.34$ \% \cite{CMD2}.}. 
We have nevertheless checked our extracted values by considering
several subsets of open channels with never more than $ \simeq 0.3 ~\sigma$
fluctuations.

\section{Conclusion}
\label{ten}

\indent
This study leads us to several conclusions. First, an expression
for the pion form factor  can be derived from the HLS Model which 
fulfills all expected analyticity requirements. In this approach,
the $\rho -\gamma$ transition amplitude becomes invariant--mass
dependent and several two--body channels couple to $\pi \pi$~;
this arises as a natural feature of the full HLS Lagrangian.
Among these additional couplings, the $\omg \pi^0$ channel
plays an interesting role as it is lower in mass than
the $K \overline{K}$ channels, more commonly accounted for.

The derived description
of timelike and spacelike experimental data is found consistent with
all the rest of the HLS phenomenology which was examined in detail
elsewhere. This includes also the HLS--KSFR relation which defines
a $\rho$ mass of $\simeq 827$ MeV perfectly accepted by 
the pion form factor data. In the present modelling, the fit
parameters are essentially subtraction constants (for the $\rho$) 
and isospin symmetry violation parameters (for the $\omg$).  

Among the
additional channels to be considered, a special role
is devoted to the $\omg \pi^0$  channel which
affects fit qualities by a significant jump in probability.
This reflects a better account of the invariant--mass region 
from the $\omg \pi^0$  threshold to the $\phi$ mass.
In contrast, the  $K \overline{K}$ channels are
found to provide no improvement and, even, no change
at all in fit qualities below the $\phi$ mass.

\vspace{0.5cm}

The model is fitted on data only sensitive to $|F_\pi(s)|$.
The phase of $F_\pi(s)$ is thus a prediction which can be
compared with  the data on the $\delta_1^1(s)$ phase shift.
It is found to match perfectly these from threshold to about
the $\rho$ mass. The agreement remains very good  
up to $ \simeq 1$ GeV and a little above independently of the
channel subset considered. All this matches well expectations
from the Watson theorem. We detect no difficulty 
which  would lead to include a $\rho(1450)$ contribution 
in order to improve the fit quality below the $\phi$ mass. 

The terms of order $s$ and $s^2$ of $F_\pi(s)$ at the chiral
point are found highly stable, with little or no sensitivity
to the inelasticity accounted for. They are found 
in fairly well  agreement with all known accepted values. 
The term of order  $s^3$ is found instead to depend sharply on 
the inelasticity accounted for~; one may question
the possibility to extract this information reliably using
only experimental data below the $\phi$ mass.

The $\omega$ branching fraction to $\pi \pi$ is found
smaller in the data set recently collected by the CMD--2 Collaboration
than in the former data sets ($2.01 \pm 0.29$ \% instead of
$2.27 \pm 0.35$ \%), however not as much as previously
claimed ($1.33 \pm 0.25$ \%).

\vspace{1.0cm}
\begin{center}
{\bf Acknowledgements}
\end{center}
Fermilab is operated by URA under DOE contract No.
DE-AC02-76CH03000.
We thank Simon Eidelman
(Budker Institute, Novosibirsk, Russia) and  Roger Forty (CERN, Geneva) 
for  discussions, comments and having  read the manuscript.
We are indebted to Wofgang Ochs (Max--Planck Institute Munich, 
Germany) for information, comments and having kindly provided 
us with the data on $\pi \pi$ phase shifts and inelasticities. 

\newpage
\appendix
\section{Loop Structure of $\Pi_{\rho \rho}(s)$ and $\Pi_{\rho \gamma}(s)$}
\label{AA}

All loops considered here should be understood amputated from
their coupling constants to external ($\gamma$  and $\rho$) lines.
As stated in the body of the text, multiparticle loops (not
present in the basic HLS Lagrangian) are not considered.

Within the non--anomalous HLS Lagrangian, the photon and the $\rho$ meson
couple both to $\pi^+ \pi^-$, $K^+ K^-$ and  $K^0 \overline{K}^0$~;
this last coupling being generated by SU(3) breaking of the ${\cal{L}}_V$
HLS Lagrangian \cite{HLS,Heath1}. Neglecting the kaon mass splitting,
this gives rise to two loop functions, given in closed form in \cite{mixing}
and named $\ell_\pi(s)$ and $\ell_K(s)$ in the body of the text.

Taking into account the anomalous (FKTUY) sector \cite{FKTUY}, other
intermediate states  have to be considered~; first, we have 
$\omega \pi^0$, $\rho^0 \eta$ and $\rho^0 \eta^\prime$, neglecting
the $\omg \phi$ mixing. This gives rise to three additional  $VP$
loops, also given in \cite{mixing}, which will be denoted
$\ell_\omg(s)$, $\ell_\eta(s)$ and $\ell_{\eta\prime}(s)$.
The couplings to $K^{*+} K^-$, $K^{*-} K^+$, $K^{*0}\overline{K}^0$,
$\overline{K}^{*0} K^0$ give rise to the same amputated loop
denoted $\ell_{K^*}(s)$, neglecting mass splittings generated
by isospin breaking. 

They come within $\Pi_{\rho \rho}(s)$ multiplied each by
the square of their coupling constants to $\rho$~; in
$\Pi_{\rho \gamma}(s)$, by the product of their coupling 
constants to $\rho$ and to the photon.
 
Whatever the (sub)set of loops effectively taken into account,
it should be stressed that this does not modify the freedom
of our model, as soon as one chooses to subtract these functions
three times~; to a large extent, these two information can
be disconnected, as one can choose externally the number of
subtractions to be performed, and there is no reason why
the number of subtractions should be minimal. 
 
Actually, increasing the subset of coupled channels 
turns out only to add definite functions with given
couplings determined numerically elsewhere by fits to radiative 
and leptonic decays. These couplings will be listed below. 
 
Taking into account the effects of $\ell_\eta(s)$, $\ell_{\eta\prime}(s)$
and $\ell_{K^*}(s)$ below the $\phi$ mass might be discussed, while
neglecting the tails of the $\rho(1450)$ and $\rho(1700)$ 
contributions or multiparticle loop effects. 
However, considering besides the pion loop,
the kaon loop with threshold at $\sqrt{s} \simeq 1 $ GeV,
while neglecting the $\omega \pi^0$ with threshold at 
$\sqrt{s} = 0.917$ GeV seems unjustified. Therefore,
we can cautiously consider that
fit results with $\pi^+ \pi^-$, $K \overline{K}$ and
$\omega \pi^0$ should be more relevant
than their analogues with only $\pi^+ \pi^-$ and $K \overline{K}$.

\section{Coupling Constants}
\label{BB}

\indent
{From} the Lagrangian piece written in Eq. (\ref{eq7}), we can derive~:

\begin{equation}
\left \{
\begin{array}{llllll}
\displaystyle  g_{\rho \pi \pi} =&\displaystyle  \frac{ag}{2} &~~,~~ &
\displaystyle g_{\gamma \pi\pi} = &\displaystyle (1-\frac{a}{2}) e \\[0.5cm]
\displaystyle g_{\rho K^+ K^-}  =& \displaystyle \frac{ag}{4z} &~~,~~ &
\displaystyle g_{\gamma K^+ K^-}= & \displaystyle (z-\frac{a}{2}-b) \frac{e}{z}\\[0.5cm]
\displaystyle g_{\rho K^0 \overline{K}^0} =&\displaystyle   -\frac{ag}{4z}&~~,~~ &
\displaystyle  g_{\gamma K^0 \overline{K}^0}= &\displaystyle  -\frac{be}{z} 
\end{array}
\right .
\label{cc1} 
\end{equation}

\noindent where $b=a(\ell_V-1)/6$. From our previous works \cite{rad,mixing,chpt},
the symmetry breaking parameters are all fixed . We have first  $z=[f_K/f_\pi]^2=3/2$
(with a remarkable precision) and $\ell_V=1.376 \pm 0.031$. We have also
obtained in these fits $a=2.51\pm 0.03$ and $g=5.65 \pm 0.02$.

{From} the anomalous Lagrangian pieces $VVP$ and $VP\gamma$ given in \cite{rad}, setting~:
\begin{equation}
\begin{array}{llll}
 \displaystyle C_\omg= -\frac{3 g^2}{8 \pi^2 f_\pi} & ~~~, ~~&  
 \displaystyle G_\omg=-\frac{3 g}{8 \pi^2 f_\pi}
\end{array}
\label{cc2} 
\end{equation}

\noindent we get~:
\begin{equation}
\left \{
\begin{array}{llllll}
g_{\rho^0 \omg \pi^0}= \displaystyle~~C_\omg & ~~~, ~~&
g_{\gamma \omg \pi^0}= \displaystyle  G_\omg~e\\[0.5cm]
g_{\rho^0 K^{*\pm} K^\mp}=\displaystyle~~ \sqrt{\frac{\ell_T}{z} }
\frac{C_\omega}{2} & ~~~, ~~&
g_{\gamma K^{*\pm} K^\mp}=\displaystyle ~~\sqrt{\frac{\ell_T}{z}}
(2-\frac{1}{\ell_T})\frac{G_\omega}{3}~e\\[0.5cm]
g_{\rho^0 K^{*0} K^0}=\displaystyle -\sqrt{\frac{\ell_T}{z}} 
\frac{C_\omega}{2} & ~~~, ~~&
g_{\gamma K^{*0} K^0}=\displaystyle -
\sqrt{\frac{\ell_T}{z}}(1+\frac{1}{\ell_T})\frac{G_\omega}{3}~e
\end{array}
\right .
\label{cc3} 
\end{equation}

\noindent with \cite{rad,mixing} $\ell_T=1.19\pm 0.06$ being
an additional breaking parameter which has been introduced
independently by \cite{Morpurgo}.
 
Defining  the physical $\eta /\eta^\prime$ fields 
in terms of singlet and octet fields $\eta_0$ and $\eta_8$
has been shown \cite{chpt}
to meet all requirements of Extended ChPT \cite{Kaiser},
including now \cite{Holstein} the extracted value for  
$\theta_0$. One could also work in the strange/non--strange
field basis \cite{Feldmann1}, but the correspondence
can be done \cite{Feldmann2} and lead to substantially
the same numerical results. Thus, defining the pseudoscalar
mixing angle by~:
\begin{equation}
\left[
     \begin{array}{ll}
     \displaystyle \eta   \\[0.5cm]
     \displaystyle \eta'   
     \end{array}
\right]
=
\left[
     \begin{array}{lll}
\displaystyle \cos{\theta_P} & -\displaystyle \sin{\theta_P} \\[0.5cm]
\displaystyle \sin{\theta_P} &
\displaystyle ~~\cos{\theta_P} 
     \end{array}
\right]
\left[
     \begin{array}{ll}
     \eta_8\\[0.5cm]
     \eta_0\\
     \end{array}
\right]
\label{cc4} 
\end{equation}
and setting $\theta_P=\theta_{ideal}+\delta_P$, we have~:
\begin{equation}
\left \{
\begin{array}{ll}
g_{\rho^0 \rho^0 \eta}= &\displaystyle \frac{C_\omega}{6}  \left[
\sqrt{2} (1-x) \cos{\delta_P} - (1+2x) \sin{\delta_P}
\right] \\[0.5cm]
g_{\rho^0 \rho^0 \eta^\prime}=&\displaystyle \frac{C_\omega}{6}  \left[
\sqrt{2} (1-x) \sin{\delta_P} + (1+2x) \cos{\delta_P}\right]\\[0.5cm]

g_{\rho^0 \gamma\eta}=&\displaystyle \frac{G_\omega}{3}  \left[
\sqrt{2} (1-x) \cos{\delta_P} - (1+2x) \sin{\delta_P}
\right] \\[0.5cm]
g_{\rho^0 \gamma \eta^\prime}=&\displaystyle \frac{G_\omega}{3}  \left[
\sqrt{2} (1-x) \sin{\delta_P} + (1+2x) \cos{\delta_P}
\right] 
\end{array}
\right . 
\label{cc5} 
\end{equation}
\noindent where \cite{chpt}  $\theta_P=-10.32^\circ \pm 0.20^\circ$.
$x$ is a parameter accounting for Nonet Symmetry breaking (no breaking
corresponding to $x=1$). It was fitted as independent parameter \cite{rad}
to $x=0.917 \pm 0.017$ with a large correlation coefficient \cite{chpt} 
$(\theta_P,x)$. In \cite{chpt}, it was shown that the observed
quasi--vanishing of $\theta_0$ implies that 
\begin{equation}
\displaystyle \theta_P=\sqrt{2} \frac{(1-z)}{2+z} ~x
\label{cc6} 
\end{equation}
\noindent is numerically well fulfilled, leading to a fit
quality identical to those obtained in \cite{rad} where this
condition was not requested~; this however lessens significantly
correlations among fit parameters.  This corresponds
to $x=0.901 \pm 0.018$, which is the value choosen for the present work.


\begin{table}
\begin{tabular}[]{|| c | p{3cm} | p{3cm} | p{3cm}  ||}
\hhhc ~~~    &        $\lambda_1$ 	 &   $\lambda_2$  &    $\lambda_3$    			\\
\hhhc ~~~    &       GeV$^{-2}$            &    GeV$^{-4}$    &    GeV$^{-6}$   		\\
\hline
ChPT \cite{Bijnens}        \hhhb &  $1.88\pm 0.07$   &  $3.85 \pm 0.60$     &  $3.0\pm 1.6$  	 \\
(without NA7)             \hhhb &  $1.88\pm 0.07$   &      $3.85 \pm 0.60$  & $4.1\pm 1.6$	\\
\hline
Ref. \cite{Truong} 	   \hhhb &    $1.93 \pm 0.06$  &  $3.90 \pm 0.20$    & $9.70\pm0.70$  	\\
\hline
Ref. \cite{Yndurain}       \hhhb&  $1.86 \pm 0.01$ & $3.60\pm0.03$  & $-$       		\\
\hline
Ref. \cite{Pich1} ($\tau$)  \hhhb&  $1.83\pm 0.05$  & $ 3.84 \pm 0.03$ & $-$          		\\
\hline
Ref. \cite{Pich1} ($e^+e^-$) \hhhb&  $1.92\pm 0.03$  & $ 3.73 \pm 0.02$ & $-$        		\\
\hline
Ref. \cite{Davier} ($\tau$)   \hhhb &  $1.89 \pm 0.04$   & $2.1 \pm 1.7$  & $15.2 \pm 5.4$           \\
\hline
Ref. \cite{Davier} ($e^+e^-$) \hhhb&  $1.89 \pm 0.04$  & $ 6.8 \pm 1.9$  & $-0.7 \pm 6.8$         \\
 \hline
 \end{tabular}
\caption{
\label{T1}
\newline
Results on the behaviour of $F_\pi(s)$ near $s=0$ from different models,
approaches and data sets. Parameters displayed are defined by Eq. (\ref{eq11}).
Entries containing the symbol $-$ are not fitted/given. }
\end{table}

\begin{table}[htb]
\begin{tabular}{|| c  | c  | c | c | c ||}
\hhhb
\hhhc ~~~    &    	~       &   ~               &   ~              &   ~~         \\
\hhhc ~~~    &  $\lambda_1  $ 	&   $\lambda_2$      &    $\lambda_3$   &  $\chi^2/dof$ \\
\hhhc ~~~    &    GeV$^{-2}$      &      GeV$^{-4}$      &    GeV$^{-6}$      &    (Prob)     \\
\hhhc   ~~   &         ~       &      ~             &   ~              & ~~            \\
\hline
\hline
$\pi^+\pi^-$ \hhhb&  $1.899\pm 0.016 $   &   $3.957\pm 0.017$     &  $10.768\pm 0.051 $  &    $183.6/178=1.03$  \\
~           \hhhb&     $- $             &      $-$               &          $-$            &  36\%\\
\hline
\hline
$\pi^+\pi^-$     $ + $ $ K\overline{K}$   \hhhb
	  &     $1.899\pm 0.016$   &   $3.958\pm 0.017$     &  $10.772\pm 0.050$  &    $184/178=1.03$    	\\
~~~~ \hhhb&     $~~$             &      $~~$               &          $ ~~$           &  36\%   \\
\hline
$\pi^+\pi^-$     $ + $ $ \omg \pi^0$   \hhhb
&  $1.899\pm 0.016$   &   $3.847\pm0.056$     &  $12.837 \pm 0.124$  &    $173.3/178$    		\\
~~~~ \hhhb&     $~~$             &      $~~$               &          $~~$            &  59\%   \\
\hline
\hline
$\pi^+\pi^-$     $ + $ $ \omg \pi^0$   \hhhb
&  $1.896 \pm 0.018$   &   $3.848 \pm 0.059$     &  $12.841\pm 0.120$  &    $173.6/178$    		\\
$ + $ $ K\overline{K}$\hhhb&     $~~$             &      $~~$               &          $~~$            &  58\%\\
\hline
\hline
$\pi^+\pi^-$     $ + $ $ \omg \pi^0$   \hhhb
&  $1.895 \pm 0.015$   &   $3.802 \pm 0.026$     &  $15.427 \pm 0.111$  &    $170.6/178=0.96$    		\\
$ + $ $ K\overline{K}$ $+$  $K^*\overline{K}$ \hhhb
&     $~~$             &      $~~$               &          $~~$            &  64\%  \\
\hline
$\pi^+\pi^-$     $ + $ $ \omg \pi^0$   \hhhb
&  $1.894 \pm 0.015$   &   $3.786 \pm 0.015$     &  $23.41 \pm 0.094$  &    $169.8/178=0.95$    		\\
$ + $ $ K\overline{K}$ $+$  $K^*\overline{K}$ \hhhb
&     $~~$             &      $~~$               &          $~~$            &  66\%  \\
$ + $ $ \rho \eta$   \hhhb
&     $~~$             &      $~~$               &          $~~$            &  $~~$  \\
\hline
$\pi^+\pi^-$     $ + $ $ \omg \pi^0$   \hhhb
&  $1.894 \pm 0.014$   &   $3.778 \pm 0.012$     &  $34.118 \pm 0.046$  &    $169.4/178=0.95$    		\\
$ + $ $ K\overline{K}$ $+$  $K^*\overline{K}$ \hhhb
&     $~~$             &      $~~$               &          $~~$            &  67\%  \\
$ + $ $ \rho \eta$ $+$  $\rho \eta^\prime$ \hhhb
&     $~~$             &      $~~$               &          $~~$            &  $~~$  \\
\hline
\end{tabular}
 
\caption{
\label{T2}
Fit results with the HLS Model. Coefficients of the
expansion of $F_\pi(s)$ near the origin~; notations
are those in Eq. (\ref{eq11}). The first column indicates
which are the coupled channels considered in the Model
function Eq. (\ref{eq4}). The number of fitted data
points is always 184, the number of free parameters is always
6, including 2 parameters for the $\omg$ contribution.  
Errors given are derived from the full error matrix computed by 
{\sc{minuit}} for the fit parameters.}
 
\end{table}


\begin{thebibliography}{99}


\bb{Yndurain}
J.~F.~De Troconiz and F.~J.~Yndurain,
``Precision determination of the pion form factor and calculation of the  muon g-2,''
Phys.\ Rev.\ D {\bf 65} (2002) 093001
[arXiv:hep-ph/0106025].


\bb{Davier}
M.~Davier, S.~Eidelman, A.~Hocker and Z.~Zhang,
``Confronting spectral functions from $e^+ e^-$ annihilation and tau decays:  
Consequences for the muon magnetic moment,''
arXiv:hep-ph/0208177.

\bb{Pich1}
A.~Pich and J.~Portoles,
``Vector form factor of the pion: A model-independent approach,''
arXiv:hep-ph/0209224.

\bb{Pich2} 
J.~J.~Sanz-Cillero and A.~Pich,
``Rho meson properties in the chiral theory framework,''
arXiv:hep-ph/0208199.

\bb{Ochs} B. Hayms {\it et al.} Nucl. Phys. {\bf B64} (1973) 134;
G. Grayer {\it et al.} Nucl. Phys. {\bf B75} (1974) 189; W. Ochs,
Doctorat Thesis, Munich 1973.

\bibitem{Aleph}
R.~Barate {\it et al.}  [ALEPH Collaboration],
``Measurement of the spectral functions of vector current hadronic tau decays,''
Z.\ Phys.\ C {\bf 76} (1997) 15.

\bb{Opal}
K.~Ackerstaff {\it et al.}  [OPAL Collaboration],
``Measurement of the strong coupling constant alpha(s) and the vector  and 
axial-vector spectral functions in hadronic tau decays,''
Eur.\ Phys.\ J.\ C {\bf 7} (1999) 571
[arXiv:hep-ex/9808019].

\bb{Protopescu} S.D. Protopopescu {\it et al.} 
``Pi Pi Partial Wave Analysis From Reactions Pi+ P 
$\to$ Pi+ Pi- Delta++ And Pi+ P $\to$ K+ K- Delta++ At 7.1-Gev/C,''
Phys. Rev. {\bf D7} (1973) 1279.

\bb{Cleo}
S.~Anderson {\it et al.}  [CLEO Collaboration],
Phys.\ Rev.\ D {\bf 61} (2000) 112002
[arXiv:hep-ex/9910046].

\bb{Ecker}
G.~Ecker, J.~Gasser, A.~Pich and E.~de Rafael,
``The Role Of Resonances In Chiral Perturbation Theory,''
Nucl.\ Phys.\ B {\bf 321} (1989) 311.

\bb{PDG02}
K.~Hagiwara {\it et al.}  [Particle Data Group Collaboration],
``Review Of Particle Physics,''
Phys.\ Rev.\ D {\bf 66} (2002) 010001.

\bb{Gounaris} G. Gounaris and J. Sakurai, 
``Finite Width Corrections To The Vector Meson Dominance Prediction For Rho $\to$ E+ E-''
Phys., Rev. Lett. {\bf 21} (1968) 244.

\bb{Barkov}
L.~M.~Barkov {\it et al.},
``Electromagnetic Pion Form-Factor In The Timelike Region,''
Nucl.\ Phys.\  {\bf B256} (1985) 365.

\bb{DM1} A. Quenzer  {\it et al.}, 
``Pion Form-Factor From 480-Mev To 1100-Mev''
Phys. Lett. {\bf B76} (1978) 512.

\bb{CMD2}
R.~R.~Akhmetshin {\it et al.}  [CMD-2 Collaboration],
Phys.\ Lett.\ B {\bf 527} (2002) 161
[arXiv:hep-ex/0112031].

\bb{Rho0} 
M.~Benayoun, S.~Eidelman, K.~Maltman, H.~B.~O'Connell, B.~Shwartz and A.~G.~Williams,
``New results in rho0 meson physics,''
Eur.\ Phys.\ J.\  {\bf C2} (1998) 269
[hep-ph/9707509].

\bb{HLS} 
M.~Bando, T.~Kugo and K.~Yamawaki,
``Nonlinear Realization And Hidden Local Symmetries,''
Phys.\ Rept.\  {\bf 164} (1988) 217.

\bb{Petersen} 
C.~D.~Froggatt and J.~L.~Petersen,
``Phase Shift Analysis Of Pi+ Pi- Scattering Between 1.0-Gev 
And 1.8-Gev Based On Fixed Momentum Transfer Analyticity. 2,''
Nucl.\ Phys.\ B {\bf 129} (1977) 89.

\bb{FKTUY} 
T.~Fujiwara, T.~Kugo, H.~Terao, S.~Uehara and K.~Yamawaki,
``Nonabelian Anomaly And Vector Mesons As Dynamical Gauge Bosons Of Hidden Local Symmetries,''
Prog.\ Theor.\ Phys.\  {\bf 73} (1985) 926.

\bb{mixing}
M.~Benayoun, L.~DelBuono, Ph.~Leruste and H.~B.~O'Connell,
``An effective approach to VMD at one loop order and the departures from  
ideal mixing for vector mesons,''
Eur.\ Phys.\ J.\ C {\bf 17} (2000) 303,
nucl-th/0004005.

\bb{Klingl} 
F.~Klingl, N.~Kaiser and W.~Weise,
``Effective Lagrangian approach to vector mesons, their structure \& decays,''
Z.\ Phys.\  {\bf A356} (1996) 193
[hep-ph/9607431].

\bb{ELOP} R.J. Eden, P.V. Landshoff, D.I. Olive J.C. Polkinhorne,
"The Analytic S--Matrix", Cambridge University Press, Cambridge UK (1966).

\bb{su2}
M.~Benayoun and H.~B.~O'Connell,
``Isospin symmetry breaking within the HLS model: A full 
($\rho$, $\omega$,  $\Phi$) mixing scheme,''
Eur.\ Phys.\ J.\ C {\bf 22} (2001) 503
[arXiv:nucl-th/0107047].

\bb{Heath1} 
M.~Benayoun and H.~B.~O'Connell,
``SU(3) breaking and hidden local symmetry,''
Phys.\ Rev.\  {\bf D58} (1998) 074006
[hep-ph/9804391].


\bb{SND}
M.~N.~Achasov {\it et al.},
``Recent results from SND detector at VEPP-2M,''
hep-ex/0010077.

\bb{rad} 
M.~Benayoun, L.~DelBuono, S.~Eidelman, V.~N.~Ivanchenko and H.~B.~O'Connell,
``Radiative decays, nonet symmetry and SU(3) breaking,''
Phys.\ Rev.\  {\bf D59} (1999) 114027
[hep-ph/9902326].


\bb{chpt}
M.~Benayoun, L.~DelBuono and H.~B.~O'Connell,
``VMD, the WZW Lagrangian and ChPT: The third mixing angle,''
Eur.\ Phys.\ J.\ C {\bf 17} (2000) 593
[arXiv:hep-ph/9905350].

\bb{BKY}
M.~Bando, T.~Kugo and K.~Yamawaki,
``On The Vector Mesons As Dynamical Gauge Bosons Of Hidden Local Symmetries,''
Nucl.\ Phys.\  {\bf B259} (1985) 493.

\bb{BGP}
A.~Bramon, A.~Grau and G.~Pancheri,
``Radiative vector meson decays in SU(3) broken effective chiral Lagrangians,''
Phys.\ Lett.\ B {\bf 344} (1995) 240.

\bb{vdm}
H.~B.~O'Connell, B.~C.~Pearce, A.~W.~Thomas and A.~G.~Williams,
``Rho - omega mixing, vector meson dominance and the pion form-factor,''
Prog.\ Part.\ Nucl.\ Phys.\  {\bf 39}, 201 (1997)
[arXiv:hep-ph/9501251].

\bb{Tony}
M.~Benayoun, H.~B.~O'Connell and A.~G.~Williams,
``Vector meson dominance and the rho meson,''
Phys.\ Rev.\ D {\bf 59} (1999) 074020
[arXiv:hep-ph/9807537].

\bb{Melikhov}
D.~Melikhov, O.~Nachtmann and T.~Paulus,
``The pion form factor at timelike momentum transfers in a dispersion  approach,''
arXiv:hep-ph/0209151.

\bb{Kaiser}
R.~Kaiser and H.~Leutwyler,
``Pseudoscalar decay constants at large N(c),''
arXiv:hep-ph/9806336.

\bb{Holstein}
J.~L.~Goity, A.~M.~Bernstein and B.~R.~Holstein,
``The decay $\pi^0 \to \gamma \gamma$ to next to leading order in chiral  
perturbation theory,''
arXiv:hep-ph/0206007.

\bb{Gardner}
S.~Gardner and H.~B.~O'Connell,
``$\rho\!-\!\omega$ mixing and the pion form factor 
in the time-like  region,''
Phys.\ Rev.\ D {\bf 57}, 2716 (1998)
[Erratum-ibid.\ D {\bf 62}, 019903 (2000)]
[arXiv:hep-ph/9707385].

\bb{GL} 
J.~Gasser and H.~Leutwyler,
``Chiral Perturbation Theory To One Loop,''
Annals Phys.\  {\bf 158} (1984) 142.
J.~Gasser and H.~Leutwyler,
``Chiral Perturbation Theory: Expansions In The Mass Of The Strange Quark,''
Nucl.\ Phys.\ B {\bf 250} (1985) 465.

\bb{Bijnens} 
J.~Bijnens, G.~Colangelo and P.~Talavera,
``The vector and scalar form factors of the pion to two loops,''
JHEP {\bf 9805} (1998) 014
[arXiv:hep-ph/9805389].

\bb{Hannah} 
T.~Hannah,
``The inverse amplitude method and chiral perturbation theory to two  loops,''
Phys.\ Rev.\ D {\bf 55} (1997) 5613
[arXiv:hep-ph/9701389].

\bb{Truong}
T. N. Truong,
``When is it possible to use perturbation technique in field theory?,''
arXiv:hep-ph/0006302.

\bb{NA7} 
S.~R.~Amendolia {\it et al.}  [NA7 Collaboration],
``A Measurement Of The Space - Like Pion Electromagnetic Form-Factor,''
Nucl.\ Phys.\ B {\bf 277} (1986) 168.


\bb{fermilab2} 
E.~B.~Dally {\it et al.},
``Elastic Scattering Measurement Of The Negative Pion Radius,''
Phys.\ Rev.\ Lett.\  {\bf 48} (1982) 375.

\bb{costa} 
B. Costa de Beauregard, T.N. Pham, B. Pire and T.N. Truong,
 ``Inelastic Effect Of The Omega Pi0 Channel On The Pion Form-Factor,''
 Phys. Lett.  {\bf B67} (1977) 213.

\bb{Morpurgo}
G.~Morpurgo,
``General Parametrization Of The V $\to$ P Gamma Meson Decays,''
Phys.\ Rev.\  {\bf D42} (1990) 1497.

\bb{Feldmann1}
T.~Feldmann,
``Quark structure of pseudoscalar mesons,''
Int.\ J.\ Mod.\ Phys.\ A {\bf 15} (2000) 159
[arXiv:hep-ph/9907491].

\bb{Feldmann2}
T.~Feldmann and P.~Kroll,
``Mixing of pseudoscalar mesons,''
Phys.\ Scripta {\bf T99} (2002) 13
[arXiv:hep-ph/0201044].

\end{thebibliography}
\end{document}